\documentclass[acmtog]{acmart}

\usepackage{booktabs} % For formal tables
\usepackage{caption}
\usepackage{subcaption}
\usepackage{microtype}
\usepackage{booktabs}
\usepackage{enumitem}
\usepackage{amsthm}
\usepackage{wrapfig}
\usepackage{multirow}
\usepackage{tabularx}
\usepackage{nicefrac}
\usepackage{dsfont}
\usepackage{amsthm}
\usepackage{overpic}
\usepackage{contour}
\contourlength{1pt}
\usepackage{xspace}
\usepackage{sidecap}
\usepackage{rotating}
\usepackage{import}
\usepackage[normalem]{ulem}

\newcommand{\kerM}{\text{ker}}
\newcommand{\imM}{\text{Image}}

\newlength{\indentlaenge}
\newlength{\mylength}
\newlength{\mylengthzwei}
\def\cput(#1,#2)#3{\put(#1,#2){\hbox to 0pt{\hss{#3}\hss}}}
\def\lput(#1,#2)#3{\put(#1,#2){\hbox to 0pt{\hss{#3}}}}
\def\rput(#1,#2)#3{\put(#1,#2){\hbox to 0pt{{#3}\hss}}}

\newcommand{\set}[1]{\mathcal{#1}}

\acmJournal{TOG}
\setcitestyle{square}
\hypersetup{draft}
\begin{document}

% Title. 
% If your title is long, consider \title[short title]{full title} - "short title" will be used for running heads.
\title{Subdivision Directional Fields}

 \author{Bram Custers}
 \orcid{0000-0001-9342-319X}
 \affiliation{
   \institution{Utrecht University/TU Eindhoven}
  }
  \email{b.a.custers@tue.nl}

 \author{Amir Vaxman}
 \orcid{0000-0001-6998-6689}
 \affiliation{
   \institution{Utrecht University}
 }
 \email{a.vaxman@uu.nl}
% abstract

% !TEX root =  SubdivisionDirectionalFields.tex

\begin{abstract}
We present a novel linear subdivision scheme for face-based tangent directional fields on triangle meshes. Our subdivision scheme is based on a novel coordinate-free representation of directional fields as halfedge-based scalar quantities, bridging the mixed finite-element representation with discrete exterior calculus. By commuting with differential operators, our subdivision is structure-preserving: it reproduces curl-free fields precisely, and reproduces divergence-free fields in the weak sense. Moreover, our subdivision scheme directly extends to directional fields with several vectors per face by working on the branched covering space. Finally, we demonstrate how our scheme can be applied to directional-field design, advection, and robust earth mover's distance computation, for efficient and robust computation.
\end{abstract}



%keywords

% A "teaser" figure, centered below the title and authors and above the body of the work.
\begin{teaserfigure}
  \centering
 \includegraphics[width=\textwidth]{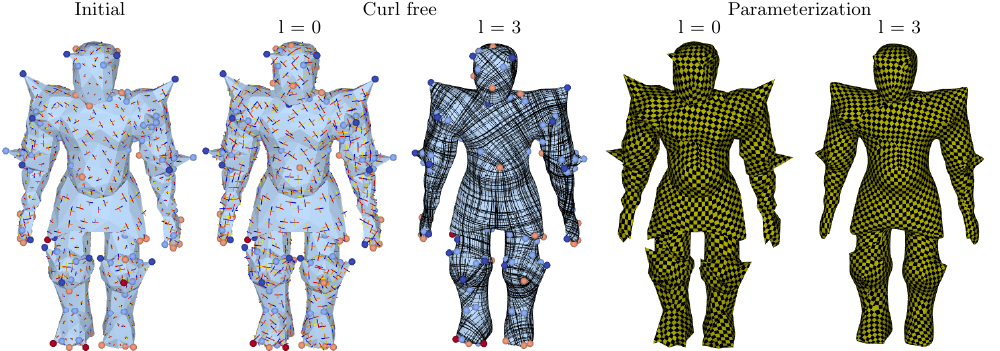}
  \caption{Rotationally-seamless parameterization with a subdivision directional field. An initial field (left) is optimized for low curl at the coarsest level $l=0$.  We subdivide the field to fine level $l=3$ (center), and then solve for a seamless parameterization in both levels (right). Our subdivision preserves curl, and thus results in a low integration error in both levels. The coarse-level optimization takes $7.5$ secs, the subdivision $7.6$ secs, and the parameterization $7.0$ secs, to a total of $22.1$ secs. This is a speedup of about two orders of magnitude compared to running the curl optimization directly on the fine level, taking $1438.7$ secs.}
  \label{fig:teaser}
\end{teaserfigure}

%% Processes all of the front-end information and starts the body of the work.
\maketitle
\newcommand{\ACc}[1]{}
%
% !TEX root =  SubdivisionDirectionalFields.tex

\section{Introduction}

Directional fields are central objects in geometry processing. They represent flows, alignments, and symmetry on discrete meshes. They are used for diverse applications such as meshing, fluid simulation, texture synthesis, architectural design, and many more. There is then great value in devising robust and reliable algorithms that design and analyze such fields. In this paper, we work with piecewise-constant tangent directional fields, defined on the faces of a triangle mesh. A directional field is the assignment of several vectors per face, where the most commonly-used fields comprise single vectors. The piecewise-constant face-based representation of directional fields is a mainstream representation within the (mixed) \emph{finite-element method} (\textbf{FEM}), where the vectors are often gradients of piecewise-linear functions spanned by values on the vertices.

Working with a fine-resolution smooth (and good-quality) mesh is often essential to get good results with methods that produce piecewise-constant directional fields. However, working on a fine mesh is also computationally expensive, and often wasteful---the desired directional fields are smooth and mostly defined by a sparse set of features such as sinks, sources, and vortices.

%Unfortunately, algorithms that are based on the piecewise-constant representation are very sensitive to the quality of the mesh and its resolution. A mesh that is too coarse or uneven would break the robustness of most applications and produce wrong results. One might alleviate such a problem by working with a fine (and good-quality) mesh.. 

A classical way to bridge this gap is to work with a multi-resolution representation, based on a nested hierarchy of meshes. A popular way to generate this representation is to use \emph{subdivision surfaces}. Subdivision surfaces are generated by operators that comprise a set of stencils, often linear and stationary (with a fixed stencil), that are used to recursively refine functions defined on meshes (and consequently the vertex positions).  These operators can be used to prolong and restrict functions between coarse and fine levels, allowing for \emph{multigrid} field computation. We consider the \emph{limit surface} as the target domain on which we compute the fields, and represent the degrees of freedom of the computation by the coarse control mesh through subdivision.

To be able to work with hierarchical directional fields on subdivision surfaces, one needs to define specialized subdivision operators. A necessary requirement for obtaining consistent results is that the subdivision operators are \emph{structure-preserving}; that is, the differential and topological properties of the directional fields are preserved under subdivision. This can be achieved by designing subdivision operators that commute with differential operators. Unfortunately, differential operators on piecewise-constant face-based fields are defined with the metric and the embedding of the mesh (e.g., face areas and normals) built in. As a result, these quantities have complicated and nonlinear expressions in the linearly-subdivided vertex coordinates. Creating linear stationary subdivision operators directly on face-based directional fields is then a challenging task. 
Recently, de Goes \emph{et al.}~\shortcite{deGoes:2016sec} devised a method for subdivision vector-field processing for differential forms in the \emph{discrete exterior calculus} (\textbf{DEC}) setting. The differential quantities in DEC are inherently separated into combinatorial and metric operators; due to this, it is possible to define a stationary subdivision scheme for differential forms that commutes with the combinatorial part alone, as introduced in~\cite{Wang:2006}.

Inspired by this insight, we introduce a coordinate-free representation for face-based fields, allowing us to decompose the face-based differential operators into independent combinatorial and metric components. With this decomposition, we define linear stationary subdivision operators for such fields. Our scheme naturally extends to branched covering spaces, where we then apply it to directional fields with an arbitrary number of vectors per face.

%\noindent In summary, our contributions are as follows:
%\begin{itemize}
%\item We introduce a coordinate-free representation for face-based directional fields, based on halfedge scalar quantities. Subsequently, we create equivalent definitions for all vector-calculus operators with this representation.
%\item We define a subdivision scheme for directional fields that commutes with the differential operators.
%\item We show how our scheme allows for efficient and robust coarse-to-fine directional field processing, where low dimensional fields on a fine mesh are spanned by fields on the coarse mesh.
%\item We demonstrate that our scheme extends to general directional fields with any number of vectors per face.
%\end{itemize}

%We demonstrate the utility and effectiveness of our subdivision scheme for directional-field design, discrete function advection,  distance computation on meshes, and seamless parameterization.

% !TEX root =  SubdivisionDirectionalFields.tex

\section{Related Work}

\subsection{Directional fields}

Tangent directional fields on discrete meshes have been researched extensively in recent years. The important aspects of their design and analysis are summarized in two relevant surveys: ~\cite{deGoes:2016vf} focuses on differential properties of mostly single vector fields, with an emphasis on different discretizations on meshes, while~\cite{vaxman:2016} focuses on discretization and representation of directional fields (with $N$ vectors at every given tangent plane) and their applications. 

The fundamental challenge of working with directional fields is how to discretize and represent them. The most common discretization considers one directional object per face, or alternatively piecewise-constant elements (e.g.,~\cite{Tong:2003,Wardetzky:2006,Bommes:2009,Crane:2010}). This representation conforms with the classic piecewise-linear paradigm of the finite-element method, and admits a dimensionality-correct cohomological structure, when mixing conforming and non-conforming elements~\cite{Wardetzky:2006}. Moreover, the natural tangent planes, as supporting plane to the triangles in the mesh, allow for simple representations of $N$-directional fields~\cite{Ray:2008,Crane:2010,Diamanti:2014}. However, the representation is only $C^0$ smooth, and makes it difficult to define discrete operators of higher order including derivatives of directional fields, such as the Lie bracket~\cite{Mullen:2011,Azencot:2013, Sageman:2019}, or Killing fields~\cite{BenChen:2010}.

An alternative approach to single-vector field processing is \emph{discrete exterior calculus}~\cite{Hirani:2003,Crane:2013}, that represents vector fields as $1$-forms, discretized as scalars on oriented edges.  DEC enjoys the benefit of representing fields in a coordinate-free manner, which allows for a decomposition of the differential operators into combinatorial and metric components. This is beneficial for the subdivision scheme we work with in this paper. However, DEC is not as of yet defined to work with general $N$-directional fields, and, when using linear Whitney forms, it still suffers from discontinuities at edges and vertices. We note that alternative approaches exist that use vertex-based definitions~\cite{Zhang:2006,Knoppel:2013,Liu:2016,Sharp:2019}, representing directional fields on tangent spaces defined at vertices. While enjoying better continuity, a full suite of differential operators has not yet been studied for them; in particular, differential operators that define discrete sequences, necessary for a correct Helmholtz-Hodge decomposition~\cite{Wardetzky:2006,Poelke:2016}.

\begin{figure*}
\includegraphics[width=0.9\textwidth]{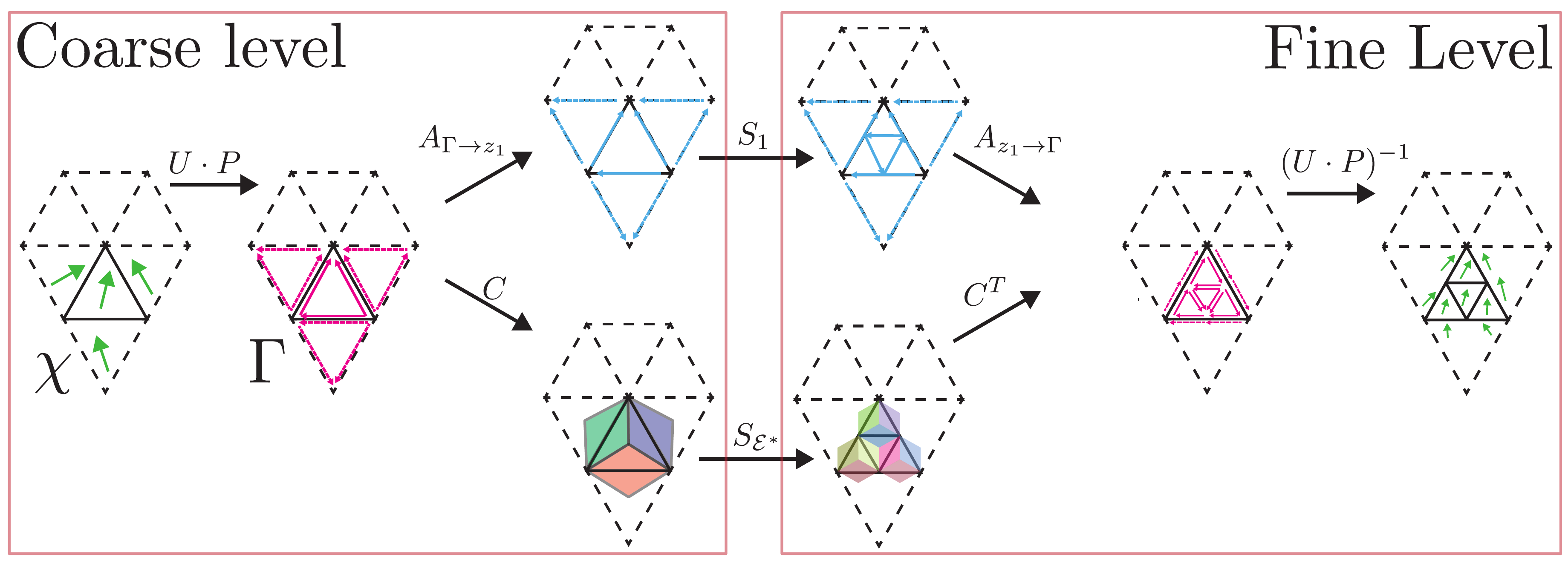}
\caption{Subdivision halfedge-form method pipeline. A face-based tangent field in the space $\mathcal{X}$ is converted to the equivalent halfedge representation in space $\Gamma$ (Section~\ref{sec:halfedge-forms}). The halfedge form is further separated into a DEC 1-form $z_1$ and a non-conforming function $\epsilon$ which is the half-curl of the field (Section~\ref{subsec:mean-curl-representation}). They are individually subdivided (Section~\ref{sec:subdivision}) and assembled back to a field on a finer mesh.}
\label{fig:subdivision-pipeline}
\end{figure*}

\subsection{Multiresolution vector calculus}

Directional fields are important for applications such as meshing~\cite{Bommes:2009,Kalberer:2007,Zadravec:2010}, simulations on surfaces~\cite{Azencot:2015}, parameterization\\~\cite{Campen:2015, Diamanti:2015, Myles:2012} and non-photorealistic rendering~\cite{Hertzmann:2000}. An underlying objective in all these applications is to obtain fields that are as smooth as possible. Nevertheless, as demonstrated in~\cite{vaxman:2016}, directional fields are subject to aliasing and noise artifacts quite easily for coarse meshes. Using fine meshes alleviates this problem to some extent, but incurs a price of increased computational overhead, especially for nonlinear methods. For this, a smooth and low-dimensional representation for smooth directional fields on fine meshes, such as the one we introduce, is much needed.

The most prevalent approach to low-dimensional smooth processing on fine meshes is to use some refinable multiresolution hierarchy. This paradigm is extensively employed in the FEM literature when using either refined elements (\emph{h-refinement}) or higher-order basis functions (\emph{p-refinement})~\cite{Babuska:1994}. This has also been applied to vector fields in planes and in volumes~\cite{Schober:2007}. A major difference in which our subdivision method departs from both these approaches is that the geometry of the target limit surface is different than that of the control coarse mesh. As such, using p- or h-refinement directly on the coarse cage is susceptible to committing the so-called ``variational crime''~\cite{Strang:2008}, where the function space and the computation domain are mismatched.

A more closesly-related prominent approach to refinable spaces is \emph{Isogeometric Analysis}~\cite{Hughes:2005}. The premise is computation over refinable B-spline basis functions, replacing the piecewise-linear FEM functions. The setting promotes integration over the target (smooth) domain, and therefore is theoretically correct and structure-preserving. However, they rely on quadrature rules to perform the complicated integrals that involve the basis functions. Methods such as~\cite{Nguyen:2014,Juettler:2016} employ subdivision rules for evaluation on the limit surface, but then design approximative quadrature rules for the exact integrals, tailored to fit specific differential operators. 

A recent work by de Goes \emph{et al.}~\shortcite{deGoes:2016sec} utilizes subdivision for $1$-forms (first introduced in~\cite{Wang:2006}) as means to represent vector fields in recursively refinable spaces. By doing so, they efficiently emulate the IGA premise in a linear setting, and directly on the discrete meshes. This technique substitutes coarse inner-product matrices with inner product matrices \emph{restricted} from the fine domains, encoding fine-mesh geometry on the coarse mesh. Using subdivision matrices as prolongation operators is akin to collapsing a single V-cycle in a multigrid setting~\cite{Brandt:1977}. The essence of the technique is to design stationary 1-form subdivision operators that commute with the discrete differential operators. This is made possible as DEC operators are purely combinatorial.

Unfortunately, their approach does not readily extend to face-based piecewise-constant fields. The effect of stationary subdivision methods on triangle areas and normals is not linear, which makes it difficult to establish the required commutation rules. Our paper introduces a novel representation of face-based fields using halfedge-based forms, that can be readily subdivided using stationary operators. As such, we introduce a metric-free subdivision method for face-based directional fields that guarantees structure preservation.

%We note that the nested spaces can be used to encode progressive details over subdivision surfaces. This is the motivation for constructing \emph{subdivision wavelets}~\cite{Lounsbery:1997,Bertram:2004} over subdivision surfaces. \BAC{Reads like we are going to do this, which we don't}

\paragraph{Directional fields} Much less has been explored in the literature about differential operators on directional fields. In~\cite{Kalberer:2007,Bommes:2009}, directional fields are used as candidate gradients for functions on branched covering spaces. Diamanti \emph{et al.}~\shortcite{Diamanti:2015} further define PolyCurl, which encodes the curl of $N$-directional fields. They then optimize for curl-free fields. However, we are not aware of any study of general directional calculus and its applications to geometry processing. We provide a branched subdivision scheme, and subsequently a multiresolution representation and a calculus suite for directional fields.

\subsection{Subdivision surfaces in geometry processing}

Subdivision surfaces are popular objects in geometry processing, and are methods of choice for shape design for animation~\cite{Liu:2014} and architectural geometry~\cite{Liu:2006}. Their most popular utility is that of multiresolution (or just coarse-to-fine) mesh editing. In the context of simulation, they have been applied to fluid simulation~\cite{Stam:2003}, thin-shell design~\cite{Cirak:2002}, and surface deformation~\cite{Grinspun:2002,Thomaszewski:2006}. The latter work also uses the folded V-cycle approach to work on the coarse mesh with the limit surface metric; nevertheless, they work with quadrature as well to approximate the exact solution.

\section{Contributions}

%\changed{Authors' comment: the entire section is new.}

The main contributions of our paper are summarized as follows.

\paragraph{Halfedge forms (Section~\ref{sec:halfedge-forms})} We define a novel coordinate-free representation for piecewise-constant vector fields on faces. The essence of this representation is to consider their projection on the halfedges defining each triangle. We prove the equivalence of this representation to that of face-based fields, and show that these \emph{halfedge forms} can be represented as the combination of a DEC $1$-form and edge-based curl, which is consistent with the case where the $1$-form is exact (the gradient of some scalar function). Halfedge forms are then a new type of $1$-form that bridges mixed-FEM representation with that of DEC. 

\paragraph{Subdivision vector fields (Section~\ref{sec:subdivision})}

Given the coordinate-free halfedge-form representation, we introduce a subdivision scheme to face-based vector fields with the following properties:

\begin{itemize}
    \item Coarse gradient fields are subdivided into fine gradient fields, where the underlying scalar function is refined using a vertex-based scalar subdivision method.
    \item The curl of a subdivided vector field, as a scalar function, is a refinement of of the curl of the coarse vector field.
\end{itemize}

We depict the subdivision pipeline in Figure~\ref{fig:subdivision-pipeline}.

\paragraph{Subdivision directional fields (Section~\ref{sec:directionals})}

Since we work with face-based fields, we show how our subdivision readily extends to $N$-directional fields, where there are $N$ vectors per face, by reducing this case into working with single-vector fields in \emph{branched} spaces.

We apply this \emph{structure-preserving} subdivision to several applications in Section~\ref{sec:applications}: earth mover's distance computation, seamless parameterization, vector field design, and operator-based advection. The common advantage that our method provides is the ability to process vector fields on subdivided meshes (with many triangles), considering only the degrees of freedom spanned by the coarse control mesh. By doing this, we save both time and memory.

We denote our face-based directional-field subdivision framework as \emph{subdivision halfedge-form method}, or in short \textbf{SHM}.

% !TEX root =  SubdivisionDirectionalFields.tex

\section{Background}
\label{sec:backgroud}

\begin{figure}
\includegraphics[width=0.35\textwidth]{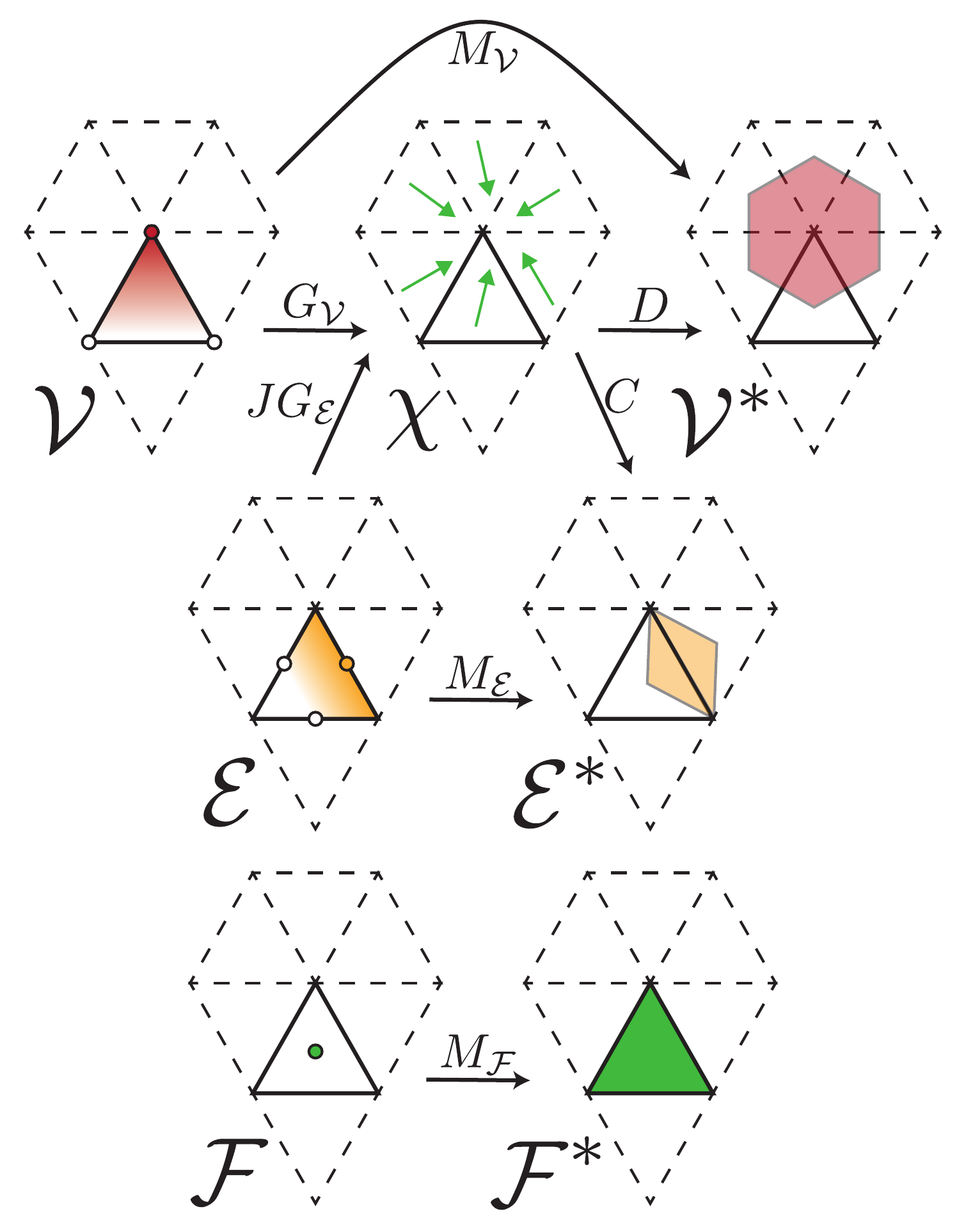}
\caption{The FEM function spaces and associated differential operators.}
\label{fig:function-spaces}
\end{figure}

We introduce a new discrete representation for vector-fields that bridges mixed FEM and DEC. For this, we require an extensive amount of background on these spaces. Nevertheless, for the sake of compactness, we mostly introduce these well-known notions in the notation and formulation we use and little else; see Table~\ref{table:notations} for our notations, Table~\ref{table:all-operators} for the definitions of the discrete differential operators, and Figure~\ref{fig:function-spaces} for the FEM space that we work in. We refer the reader to~\cite{Wardetzky:2006,deGoes:2016vf} for a more comprehensive account of the operators in FEM, and to~\cite{Desbrun:2005} for the operators in DEC. For compactness, we reduce the polysemous ``FEM'' to only mean the conforming/non-conforming piecewise-linear finite element representation, in order to distinguish it from DEC, which is in essence another type of finite-element representation.

\subsection{Function spaces}
\label{subsec:function-spaces}

\begin{table*}[h]
\centering
\begin{tabularx}{\textwidth}{|c|c|X|}
\hline 
\textbf{Notation} & \textbf{Dimensions} &\textbf{Explanation} \\
\hline
$\mathcal{V}, \mathcal{V}^{\ast}$ & $\left|V\right|$ & Primal and dual vertex-based PL conforming functions. \\
$\mathcal{E}, \mathcal{E}^{\ast}$ & $\left|E\right|$ & Primal and dual midedge-based PL non-conforming functions. \\
$\mathcal{F}, \mathcal{F}^{\ast}$ & $\left|F\right|$ & Primal and dual face-based PC functions. \\
$\mathcal{X} $& $3\left|F\right|$ & Piecewise-constant face-based vector fields (PCVFs).\\
$\mathcal{Z}_1 $& $\left|E\right|$ & Edge-based DEC 1-forms.\\
$\Gamma$ & $2\left|F\right|$ & Halfedge forms.\\
$P$ & $\left|\Gamma\right| \times \left|\mathcal{X}\right|$ & Projection operator $\mathcal{X} \rightarrow \Gamma$ (Eq.~\ref{eq:packed-projection}). \\
$U$ & $2\left|F\right| \times 3\left|F\right|$ & Unpacking operator for halfedge form in each face, respecting null-sum (Eq.~\ref{eq:packing-operator}).\\
$M_\mathcal{V}$ & $\left|V\right| \times \left| V\right|$ & Mass matrix for vertices (Voronoi areas). \\
$M_\mathcal{E^{\ast}}$ & $\left|E\right| \times \left| E\right|$ & Mass matrix for integrated edge quantities (\emph{inverse} diamond areas). \\
$M_\mathcal{F}$ & $\left|F\right| \times \left| F\right|$ & Mass matrix for face (triangle areas). \\
$M_\Gamma$ & $2\left|F\right| \times 2\left| F\right|$ & Mass matrix for halfedge forms (packed cotangent weights; Eq.~\ref{eq:m-gamma}). \\
$M_\mathcal{X}$ & $3\left|F\right| \times 3\left| F\right|$ & Mass matrix for PCVFs. This amounts to repeating $M_\mathcal{F}$ three times per triangle. \\
$S_\mathcal{P}^l$ & $\left|\mathcal{P}^{l+1}\right| \times \left|\mathcal{P}^l\right|$ & Subdivision matrix for space $\mathcal{P}$ from level $l$ to $l+1$. We use $\mathcal{P} \in \left\{\mathcal{V}, \mathcal{E^{\ast}}, \mathcal{F^{\ast}}, \mathcal{Z}_1, \Gamma\right\}$. \\
$\mathbb{S}_\mathcal{P}^l$ & $\left|\mathcal{P}^{l}\right| \times \left|\mathcal{P}^0\right|$ & Aggregated subdivision matrix from levels $0$ to $l$.\\
$\mathbb{M}^0_\mathcal{P}$ & $\left|\mathcal{P}^{0}\right| \times \left|\mathcal{P}^0\right|$ & Restricted mass matrix from some level $l$ to level $0$ (Eq.~\ref{eq:mass-restriction}).\\
$A_{\mathcal{Z}_1 \rightarrow \Gamma}$ & $\left|\Gamma\right| \times \left| E\right|$ & Assigning a $1$-form value from an edge to its halfedges. $A_{\Gamma \rightarrow \mathcal{Z}_1}=\left(A_{\mathcal{Z}_1 \rightarrow \Gamma}\right)^T$ sums the halfedge values to a single $1$-form per edge.\\
$A_{\mathcal{E}^{\ast} \rightarrow \mathcal{F^{\ast}}}$ & $\left|F\right| \times \left| E\right|$ & Summing integrated edge quantities to the adjacent faces.\\
$W$ & $\left|\Gamma\right| \times 2\left|E\right|$ & Computing edge mean $1$-form $z_1$ and half-curl $\epsilon$ (Eq.~\ref{eq:W}).\\
$\mathcal{P}^N$ & $\left|\mathcal{P}\right|^N$ & $N$-branched space of functions in $\mathcal{P}$. e.g., $N$-directional fields are in $\mathcal{X}^N$  \\
\hline
\end{tabularx}
\caption{List of notations and symbols. For differential operators see Table~\ref{table:all-operators}. Superscript capital $N$ is for number of vectors and small $l$ is for subdivision level.}
\label{table:notations}
\end{table*}

We work with a triangle mesh $\set{M} = \left(V,E,F\right)$ of arbitrary genus, and with or without boundaries. As we combine FEM and DEC formulations, we need to streamline notation at the expense of conventionality. We define $\mathcal{V}$ as the space of piecewise-linear (conforming) vertex-based functions, corresponding to 0-forms with linear Whitney forms in DEC and $\mathcal{S}_h$ in FEM. We further define $\mathcal{E}$ as the space of piecewise-linear mid-edge (non-conforming) functions, also known as the \emph{Crouzeix-Raviart} elements~\cite{Crouzeix:1973}, corresponding to $\mathcal{S}^{\ast}_h$ in FEM. We define $\mathcal{F}$ as the space of piecewise-constant functions on faces, corresponding with dual 2-forms in DEC.  We define the corresponding integrated (weak) function spaces on vertices as $\mathcal{V}^{\ast}$ (corresponding to dual 0-forms, integrated over Voronoi areas), on edges as $\mathcal{E}^{\ast}$ (integrated over edge diamond areas), and on faces as $\mathcal{F}^{\ast}$ (corresponding to primal 2-forms in DEC). Finally, we denote the space of \emph{face-based piecewise-constant directional fields} (\textbf{PCDF}) of degree $N$, defined on the tangent spaces spanned by the supporting planes to the faces, as $\mathcal{X}^N$. The latter is in accordance with the conventional notation. We introduce our operators to the classic case of $N=1$, and then generalize our constructions to $N$-directional fields in Section~\ref{sec:directionals}. For case $N=1$, we omit the power and just use $\mathcal{X}$, the space of piecewise-constant vector fields (\textbf{PCVF}).

\begin{wrapfigure}{r}{0.25\textwidth}
\vspace{-0.2In}
    \includegraphics[width=0.25\textwidth]{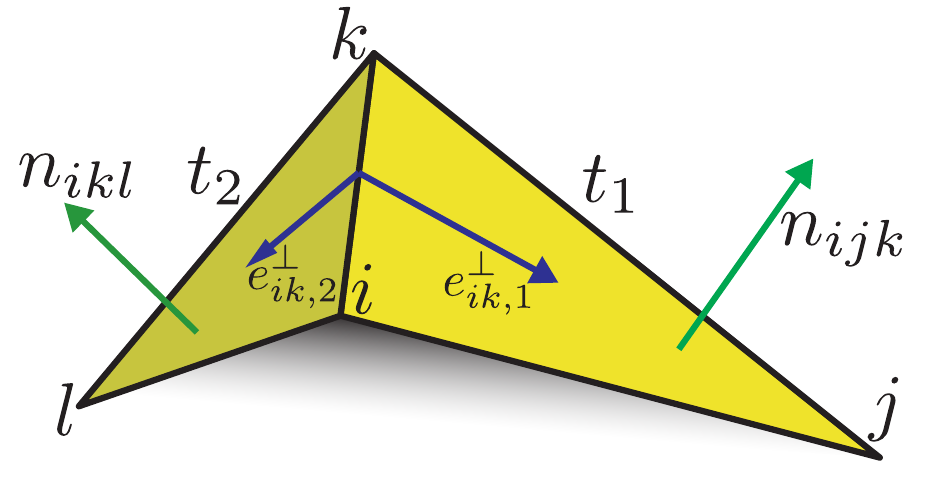}
  \caption{Our notation for a single flap.}
  \label{fig:single-flap}
\end{wrapfigure}

\paragraph{Orientation}
We choose an arbitrary (but fixed) orientation for every edge in the mesh. This orientation consistently defines both source and target vertices (\emph{primal orientation}), and left and right faces for each edge (\emph{dual orientation}; corresponding with the CCW orientation of every face). For instance, in our notation, we use $e_{ik}$ and get $\textit{left}(e_{ik}) = ikl = t_2$ and $\textit{right}(e_{ik}) = ijk = t_1$ (see Fig.~\ref{fig:single-flap}). For edge $e$ and adjacent face $f$, we define $s_{e,f}=\pm 1$ as the sign encoding the orientation (positive if $f=left(e)$, i.e. $e$ is oriented CCW with respect to the face normal of $f$). DEC $1$-forms depend on the direction and sign of the edge, so they are denoted as \emph{oriented quantities}. Quantities in $\mathcal{E}^{\ast}$ depend on the direction of the edge on which they are defined, but not on the specific sign (whether $e_{ik}$ or $e_{ki}$), and thus we denote them as \emph{unsigned quantities}.  For a face $f$, we define $J_{|f}=\left[\hat{n}_f \times \right]$ as the operator that performs the rotation around its normal $\hat{n}_f$.

\paragraph{Mixing spaces} It is well-known~\cite{Polthier:2003} that the discrete differential FEM operators preserve the structure of differential operators in the discrete setting. That is, we have a sequence: $\imM(G_\mathcal{V}) \subset \kerM(C_\mathcal{E^{\ast}})$ (gradient fields are curl-free) and a (dual) sequence: $\imM(JG_\mathcal{E}) \subset \kerM(D_\mathcal{V})$ (rotated \emph{cogradient} fields are divergence free). This \emph{structure-preserving} property is essential to the correct and stable behaviour of differential equations discretized with such operators. Note that the entire formulation can be done in a dual manner by switching conforming and non-conforming spaces and operators. However, we restrict ourselves to conforming gradients and non-conforming rotated cogradients. As such, we omit the space-indicating subscripts and just use $D$ for (conforming) divergence and $C$ for (non-conforming) curl.

\paragraph{Helmholtz-Hodge decomposition}
Mixing conforming and non-conforming operators is essential to have a dimensionality-consistent Hodge decomposition~\cite{Wardetzky:2006}. For a closed surface without boundary, there is a well-defined Helmholtz-Hodge decomposition of $\mathcal{X}$ as follows:
\begin{equation}
\mathcal{X} = \imM\left(G_\mathcal{V}\right) \oplus \imM\left(JG_\mathcal{E}\right) \oplus \mathcal{H}_\mathcal{X}.
\end{equation}
$\imM(G_\mathcal{V})$ is the space of vectors fields that are gradients of functions in $\mathcal{V}$, $\imM(JG_\mathcal{E})$ is the space of rotated cogradients of functions in $\mathcal{E}$, and $\mathcal{H}_\mathcal{X}=\text{ker}(C) \bigcap \kerM(D)$ is the space of PCVF \emph{harmonic fields}. The space of harmonic fields has the correct dimension $2g$, where $g$ is the genus of the mesh. 

\paragraph{Inner products} Inner products on the function spaces are represented as mass matrices $M$, where two elements $u,v$ in column vector form have the inner product $\langle u, v \rangle_\mathcal{P} = u^T M_\mathcal{P} v$ in some function space $\mathcal{P}$. $M_\mathcal{X}$ is the mass matrix of space $\mathcal{X}$, comprising diagonal values of triangle areas for each component of the vector field, and we further define $M_\mathcal{V}$ to be the diagonal matrix of \emph{Voronoi areas} of every vertex. We define $M_\mathcal{E}$ to be the diagonal matrix of \emph{diamond areas} supported on each edge (See Fig.~\ref{fig:function-spaces}). Mass matrices for dual spaces are inverted mass matrices of the corresponding primal spaces. We note that $M_\mathcal{V}$ and $M_\mathcal{E}$ are in fact \emph{lumped} versions of the FEM mass matrices. This lumping is done to make them diagonal, and thus have simple inverses. We denote the $L_2$ norm of space $\mathcal{P}$ by $|| u ||_\mathcal{P} = \sqrt{\langle u, u \rangle_\mathcal{P}}$.

\paragraph{Hodge Laplacian} 

%The well-known integrated conforming Laplacian operator $L_\mathcal{V}:\mathcal{V}\rightarrow \mathcal{V}^{\ast}$ is defined in FEM as $L_\mathcal{V} = D\cdot G_\mathcal{V} = G_\mathcal{V}^T \cdot M_\mathcal{X} \cdot G_\mathcal{V}$. Note that it reveals the nature of the Laplacian matrix as an inner-product matrix of the form $\langle \nabla f_1, \nabla f_2 \rangle$ for vertex-based functions $f_1,f_2 \in \mathcal{V}$. The result is the \emph{cotangent Laplacian}:

%\begin{align}
%\left( L_\mathcal{V} \right)_{ik} = \left\{ \begin{array}{lr} i=k & %\sum_{\left(i,k\right) \in E} {w_{ik}} \\ i \neq k & -w_{ik} \end{array} \right\},
%\label{eq:conforming-laplacian}
%\end{align}

%where $w_{ik}=\frac{cot(\angle{ijk})+cot(\angle(kli))}{2}$. The integrated non-conforming Laplacian $L_\mathcal{E}$ is consequently
%$$L_\mathcal{E}=G_\mathcal{E}^T\cdot J^T \cdot M_\mathcal{X}\cdot J\cdot G_\mathcal{E} = G_\mathcal{E}^T\cdot M_\mathcal{X}\cdot G_\mathcal{E}.$$

%The pointwise versions of these Laplacians are $\left(M_\mathcal{V}\right)^{-1}L_\mathcal{V}$ and $\left(M_\mathcal{E}\right)^{-1}L_\mathcal{E}$.

The integrated discrete Hodge Laplacian is obtained from minimizing the Dirichlet energy of vector fields, and has the following form~\cite{Brandt:2016}: $$L_\mathcal{X} = C^TM_{\mathcal{E}^{\ast}}C + D^TM_{\mathcal{V}^{\ast}}D.$$ Its null-space contains the harmonic vector fields. The pointwise version is $M_\mathcal{X}^{-1} L_\mathcal{X}$.

\renewcommand{\arraystretch}{1.5}
\begin{table*}[h]
	\centering
	\begin{tabular}{|c|c|c|c|c|c|c|}
		\hline
		\textbf{Operator}& \multicolumn{2}{c|}{\textbf{FEM}} & \multicolumn{2}{c|}{\textbf{ DEC}} & \multicolumn{2}{c|}{\textbf{$\Gamma$}} \\
		\hline
		 & \textbf{Spaces} & \textbf{formulation} & \textbf{Spaces} & \textbf{formulation} & \textbf{Spaces} & \textbf{formulation} \\ 
			\hline
		Primal gradient & $\mathcal{V} \rightarrow \mathcal{X}$ & $G_\mathcal{V}$ & $\mathcal{V} \rightarrow \mathcal{Z}_1$ & $d_0$ & $\mathcal{V} \rightarrow \Gamma$ & $d_{0,\Gamma} = U^{-1} A_{\mathcal{Z}_1 \rightarrow \Gamma} d_0$\\
				\hline
		Dual rotated gradient & $\mathcal{E} \rightarrow \mathcal{X} $ & $JG_\mathcal{E}$ & $\mathcal{F^{\ast}} \rightarrow \mathcal{Z}_1$ & $(M_1)^{-1} d_1^T$ & $\mathcal{E} \rightarrow \Gamma$ & $-(M_\Gamma)^{-1} C_\Gamma^T$\\
				\hline
		Divergence & $\mathcal{X} \rightarrow \mathcal{V}^{\ast} $ & $D = G_\mathcal{V}^{T} M_\mathcal{X}$ & $\mathcal{Z}_1 \rightarrow \mathcal{V^{\ast}}$  & $d_0^T M_1$ &$\Gamma \rightarrow \mathcal{V^{\ast}}$ & $D_\Gamma=\left(d_{0,\Gamma}\right)^T M_{\Gamma}$\\
				\hline
		Curl & $\mathcal{X} \rightarrow \mathcal{E^{\ast}}$ & $C$ & $\mathcal{Z}_1 \rightarrow \mathcal{F^{\ast}}$ &  $d_1$ & $\Gamma \rightarrow \mathcal{V^{\ast}}$& $C_\Gamma = C\cdot U$\\
				\hline
		Primal Laplacian & $\mathcal{V} \rightarrow \mathcal{V^{\ast}}$ & $L_\mathcal{V} = G_\mathcal{V}^T M_\mathcal{X} G_\mathcal{V}$ & $\mathcal{V} \rightarrow \mathcal{V^{\ast}}$ & $d_0^T M_1 d_0$ & $\mathcal{V} \rightarrow \mathcal{V^{\ast}}$ & $\left(d_{0,\Gamma}\right)^TM_\Gamma d_{0,\Gamma}$\\
				\hline
		Dual Laplacian & $\mathcal{E} \rightarrow \mathcal{E^{\ast}}$ & $L_\mathcal{E} = (JG_\mathcal{E})^T M_\mathcal{X} JG_\mathcal{E}$ & $\mathcal{E} \rightarrow \mathcal{E^{\ast}}$ & $d_1 M_1^{-1} d_1^T$ & $\mathcal{E} \rightarrow \mathcal{E^{\ast}}$ & $ C_{\Gamma} M_\Gamma^{-1} \left(C_{\Gamma}\right)^T$\\
				\hline
		Hodge Laplacian & $\mathcal{X}\rightarrow \mathcal{X}$ & $L_\mathcal{X} = G_\mathcal{V} M_{\mathcal{V}^{\ast}} G_\mathcal{V}^T M_\mathcal{X} +$ & $\mathcal{Z}_1 \rightarrow \mathcal{Z}_1$ & $L_1 = d_0 M_{\mathcal{V}^{\ast}} d_0^T M_1 +$  & $\Gamma \rightarrow \Gamma$  & $ L_\Gamma = d_{0,\Gamma} M_{\mathcal{V}^{\ast}}\left(d_{0,\Gamma}\right)^T \cdot M_\Gamma+$ \\
		& & $J G_\mathcal{E} M_{\mathcal{E}^{\ast}} (JG_\mathcal{E})^T M_\mathcal{X}$ & & $M_1 ^{-1}d_1^T {M_\mathcal{F}^{\ast}} d_1$  & & $ M_\Gamma^{-1}\left(C_{\Gamma}\right)^T\ M_{\mathcal{E}^{\ast}} C_{\Gamma}$\\
		\hline
	\end{tabular}
\caption{Operators per representations. All operator are presented in their integrated versions when applicable.}
\label{table:all-operators}
\end{table*}

\subsection{Discrete Exterior Calculus}
\label{subsec:dec}

\paragraph{DEC function spaces} The setup of DEC~\cite{Desbrun:2005} on surface meshes is an alternative to the piecewise-constant representation. Instead of representing vectors explicitly, DEC works with primal and dual $k$-forms, where primal $0$-forms are (pointwise) vertex-based functions, primal $1$-forms are (integrated) edge-based functions (representing vectors), and primal $2$-forms are (integrated) face based functions. The space of primal $0$ forms $\mathcal{Z}_0$, with the interpolation of linear \emph{Whitney forms}, identifies with $\mathcal{V}$. The space of $1$-forms $\mathcal{Z}_1$ comprises scalars on edges, representing oriented quantities. Such quantities are oriented in the sense that when a scalar $z$ is attached to edge $e_{ik}$, then the corresponding scalar for the edge $e_{ki}$ is $-z$. Note that the FEM space $\mathcal{E}^{\ast}$ does not have this property or edge sign dependence and therefore it does not identify with $\mathcal{Z}_1$. The space of $2$-forms $\mathcal{Z}_2$ identifies with $\mathcal{F}^{\ast}$ (note the dual space, as elements in $\mathcal{Z}_2$ are integrated).

The space of dual $0$-forms $\mathcal{Z}^{\ast}_0$ are integrated vertex-based quantities, and identifies with $\mathcal{V}^{\ast}$. Similarly, $\mathcal{Z}_2^{\ast}$ identifies with $\mathcal{F}$. Dual $1$-forms in the space $\mathcal{Z}_1^{\ast}$ are defined on the union of the orthogonal duals to the edges. For edge $ik$ in triangles $ijk$ and $ikl$, the dual $e^{\ast}_{ik}$ is the two perpendicular bisectors to $e_{ik}$ from the center of the circumscribing circles of each triangle, and therefore differs from the rotated edge $e_{ik}^{\perp}$ used in FEM.

\paragraph{Differential operators} Two fundamental discrete operators are combined to create an entire suite of vector calculus: the exterior derivative $d$, taking $k$-forms into $\left(k+1\right)$-forms, and the Hodge star $\star$, taking primal $k$-forms into dual $2-k$ dual forms. For instance, the lumped $\star_1: \mathcal{Z}_1 \rightarrow \mathcal{Z}_1^{\ast}$ is defined as $\star_{|ik,1} = \frac{\left|e^{\ast}_{ik}\right|}{\left|e_{ik}\right|}$. To streamline notation, we use $M_1$ to represent $\star_1$. $M_1:\left|E\right| \times \left|E\right|$ is a diagonal matrix that contains the weights per edge. $M_0$ identifies with $M_\mathcal{V}$, as a diagonal matrix of Voronoi areas, and $M_2$ identifies with $M_{\mathcal{F}^{\ast}}$.

DEC operators also define a (de-Rham) sequence, as $d^2=0$ in the discrete setting. Therefore DEC is also structure preserving. In the dual setting, we also work with the boundary operator $\partial = d^T$. Intuitively, $\partial$ sums up $\left(k+1\right)$-forms into $k$-forms of elements (chains) adjacent to them, with relation to the mutual orientation. The vector calculus operators are then interpreted as follows: the curl operator is simply $d_1$, where curl is a primal $2$-form in DEC, and primal (weak) divergence is $\left(d_0\right)^TM_1$, producing a dual $0$-form.  

%Curl-free $1$-forms, where $d_1z_1=0$ are called \emph{closed}, and $1$-forms $z_1$ for which there exists a $0$-form $z_0$ so that $z_1 = d_0z_0$ are called \emph{exact} (paralleling the notion of gradient fields in FEM). It is evident that exact forms are by definition closed. Divergence-free $1$-forms, where $d_0^T M_1 z_1 =0$ are called coclosed, and $1$-forms for which there exists a $2$-form such that $z_1 = M_1^{-1}\left(d_1\right)^TM_2z_2$ are called \emph{coexact} (paralleling rotated cogradient fields in FEM). 

The DEC version of Hodge decomposition for $1$-forms is such that for each $z_1 \in \mathcal{Z}_1$ there exist $z_0 \in \mathcal{Z}_0$ and $z_2 \in \mathcal{Z}_2$ such that:
\begin{equation}
z_1 = d_0z_0 + M_1^{-1}d_1^TM_2z_2 + h_1.
\end{equation}
where $h_1$ is a harmonic $1$-form that is both closed and coclosed.

\paragraph{Between DEC and FEM} As linear discrete frameworks, DEC and FEM admit a similar power of expression, for instance $L_0 = L_\mathcal{V}$, the cotangent weights Laplacian. However, they are incompatible otherwise; $\left|\mathcal{Z}_1\right|=\left|E\right|$, while $\left|\mathcal{X}\right|=2\left|F\right|$ (the ambient dimension in the raw representation is $3\left|F\right|$). As such, the differential operators are also different in dimensions. 

Note that the commonly used diagonal $M_1$ is a lumped version of the ``correct'' (Galerkin) mass matrix for $1$-forms, integrating over the interpolated linear Whitney forms~\cite{deGoes:2014}. The lumped version results in diagonal matrices that are efficient to work with, especially with regards to solving equations. Moreover, interpolated closed (and, as a subset, exact) $1$-forms are piecewise-constant; in that case, the lumped $M_1$ is the correct inner product. This is the reason that FEM and DEC vertex Laplacians identify.

DEC has an advantage over FEM in the sense that it allows for a natural separation between the combinatorial differential operator $d$, and the metric encoded in the mass matrices, whereas PCVF spaces do not exhibit this separation. This distinction plays an important part in our definition of the subdivision operators. 

\subsection{Subdivision Exterior Calculus}
\label{subsec:subdivision-exterior-calculus}

\paragraph{Subdivision surfaces} A subdivision surface is a hierarchy of refined meshes, starting from a coarse \emph{control mesh}, and converging into a smooth fine mesh. We focus on approximative triangle-mesh schemes for both vertex-based and face-based functions. Extending notation from~\cite{Wang:2006,deGoes:2016sec}, we denote a subdivision operator as $S_{\mathcal{P}}^{l}$, where it subdivides an object of space $\mathcal{P}$ defined on a mesh in level $l$, denoted as $\set{M}^l$, to an object on a mesh of the refined space in $\set{M}^{l+1}$. For instance, $S_{\mathcal{E}^{\ast}}^5$ subdivides an unsigned integrated edge quantity in $\mathcal{E}^{\ast}$ from level $5$ to level $6$. 

We denote the product of subdivision matrices from the coarsest level to a given level $l$ as: $\mathbb{S}_\mathcal{P}^{l} = \prod_{i=0}^{l-1} {S_{\mathcal{P}}^i}$. The columns of $\mathbb{S}_{\mathcal{P}}^l$ converge into refined basis functions $\Psi_{\mathcal{P}}^{0}$ defined on $\mathcal{M}^l$. These basis functions admit a nested refinable heirarchy:
\begin{equation}
\Psi_{\mathcal{P}}^0 \subset \Psi_{\mathcal{P}}^1 \subset \cdots \subset \Psi_{\mathcal{P}}^l,
\end{equation}
where a function $\Psi_{\mathcal{P}}^{k}$ 
is a linear combination of basis functions at level $\Psi_{\mathcal{P}}^{k+1}$, encoded in the matrix $S_{\mathcal{P}}^k$: $\Psi_{\mathcal{P}}^k = \Psi_{\mathcal{P}}^{k+1}S_{\mathcal{P}}^k$. Note that $\Psi_\mathcal{P}^0 = \Psi_\mathcal{P}^l \mathbb{S}^l_\mathcal{P}$.

\paragraph{Structure-preserving subdivision}

The essence of Subdivision Exterior Calculus (\textbf{SEC}) is the definition of stationary subdivision matrices for $k$-forms that commute with the differential as follows:
\begin{align}
\nonumber d_0 S_0 &= S_1d_0, \\ 
d_1 S_1 &= S_2d_1.
\end{align}
This commutation subdivides exact $1$-forms into exact $1$-forms where the underlying $0$-form is refined. Similarly, the curl of a fine $1$-form is the subdivided curl of the coarse $1$-form.

\paragraph{Restricted inner products} 
Choosing Loop subdivision~\cite{Loop:1987,Biermann:2000} for $S_0$ and half-box spline subdivision~\cite{Prautzsch:2002} for $S_2$ completely defines $S_1$, with some assumptions on the symmetry of the $S_1$ stencil. In~\cite{deGoes:2016sec}, the subdivision operator is mainly used for the purpose of defining mass matrices on the coarse mesh as \emph{restricted} fine mass matrices: 
\begin{equation}
\mathbb{M}^0_\mathcal{P} = \left(\mathbb{S}^l_\mathcal{P}\right)^T \cdot M^l_\mathcal{P} \cdot \mathbb{S}^l_\mathcal{P}
\label{eq:mass-restriction}
\end{equation}
for the space $\mathcal{P}$ and associated subdivision matrix $\mathbb{S}^l_\mathcal{P}$ from level $0$ to level $l$ as above.
The restricted mass matrix $\mathbb{M}^0_\mathcal{P}$ is exactly the product between subdivided $\mathcal{P}$-forms in the fine level $l$. The restricted mass matrices are in general no longer diagonal; however, they have a limited support (usually just two rings), derived from the support of the subdivision matrix. Working with restricted mass matrices provides SEC with a smooth, localized, and small function space on the limit surface in a structure-preserving manner that does not require special treatment for singularities, replacing the quadrature methods employed by IGA.

%An advantage of the restricted mass matrices is improving the conditioning of the operators for which they are building blocks; this is the case since stationary subdivision operators, as uniform averaging operators, improve the quality of the mesh in the finer levels.

\paragraph{Divergence pollution} The relation of the SEC divergence to the fine DEC divergence reveals an interesting insight:
\begin{align}
\label{eqn:sec-div-kernel}
\left(d_0\right)^T\mathbb{M}^0_1z^0_1 &= \left(d_0\right)^T \left(\mathbb{S}^l_1\right)^T \cdot M^l_1 \cdot \mathbb{S}^l_1 z^0_1 \\ \nonumber
&= \left(\mathbb{S}^l_0\right)^T \left(d_0\right)^T \cdot M^l_1 \cdot \left(\mathbb{S}^l_1 z^0_1\right) \\ \nonumber
&= \left(\mathbb{S}^l_0\right)^T \left(\left(d_0\right)^T \cdot M^l_1 \cdot z^l_1\right). \\ \nonumber
\end{align}
In words, the SEC divergence of a coarse $1$-form $z_1^0$ subdivided into fine $1$-form $z^l_1$ is  not exactly the subdivided coarse divergence; it is rather equivalent only when tested against the \emph{test functions} $\Psi_0^0$. Simply put, the divergence of the fine form might contain ``high-frequency'' components that are in $\ker\left(\mathbb{S}^l_1\right)^T$, where $\left(\mathbb{S}^l_1\right)^T$ acts effectively as a low-pass filter. We denote this as \emph{divergence pollution}.

%The framework of SEC does not trivially extend to FEM; this is because FEM operators do not factor into combinatorial and metric components, and thus creating stationary subdivision matrices for PCVFs is a challenging task \changed{that we handle in this work.}

% !TEX root =  SubdivisionDirectionalFields.tex

\section{Halfedge Forms}
\label{sec:halfedge-forms}

%As we mention in the previous section, FE vector-field spaces do not enjoy this separation, and it is therefore impossible to define stationary subdivision rules that commute with FE differential operators directly; any subdivision would have to contend with the metric changes (for instance, the change in tangent planes with the way normals behave under subdivision). We would like to preserve the ability to use stationary rules within this space. 

We aim to create a stationary subdivision scheme for PCVFs, inspired by SEC, achieving our goal of establishing a framework of hierarchical spaces for directional fields. For this, we need to first overcome the challenge of metric-free representation that allows for stationary commutation. We do so in the following by creating a halfedge representation for $\mathcal{X}$.

For each oriented edge $e_{ik}$ adjacent to faces $t_1$ and $t_2$, we consider its \emph{halfedges} $e_{ik,1}$ and $e_{ik,2}$ (with the notation of Fig.~\ref{fig:single-flap}). Note that they are both oriented in the same direction as $e_{ik}$; this departs from the usual doubly-connected edge list convention~\cite{deBerg:2008}, where halfedges are of opposing orientations, and counterclockwise oriented with respect to their face normal. We choose to co-orient them with the edge as it is a more natural convention for our differential operators. 

We define $\Gamma$ as the space of \emph{null-sum} oriented scalar quantities on halfedges: for every face $t$ with halfedges $e_{1,t}, e_{2,t}, e_{3,t}$, and with signs $s_{1,t},s_{2,t},s_{3,t}$ that encode the orientation of the respective halfedges with regards to $t$ (see Section~\ref{subsec:function-spaces}), we consider corresponding scalar quantities $\gamma_{1,t}, \gamma_{2,t}, \gamma_{3,t}$ that must satisfy: 
\begin{equation}
s_{1,t}\gamma_{1,t}+s_{2,t}\gamma_{2,t}+s_{3,t}\gamma_{3,t}=0.
\end{equation}
We denote $\gamma = \left\{\gamma_{e,t}\right\} \in \Gamma$ as a \emph{halfedge form}.

\paragraph{Equivalence to $\mathcal{X}$} We represent the halfedges as row vectors $e_{.,t}$. With that, we define the projection operator $P':\mathcal{X} \rightarrow \Gamma$ as follows:
\begin{equation}
P'_{|t}=\begin{pmatrix} e_{1,t} \\ e_{2,t} \\ e_{3,t} \end{pmatrix}.
\end{equation}
Note that $P'_{|t}$ has zero row sum, which is the sum of edges of a single triangle oriented with the proper signs; its null space is spanned by vectors along the normal of the triangle. For each $v \in \mathcal{X}$, the null sum of $\gamma = P'v$ is trivially satisfied. The operator $P'$ is analogous to the ``$\flat$'' operator that converts a vector field to a $1$-form.

Conversely, for every $\gamma \in \Gamma$, which has null sum by definition, the system $P'v=\gamma$ has a single solution that is also a tangent vector (without normal components)---it can be reproduced by the Penrose-Moore pseudo-inverse $v = P'^{-1}\gamma$ (the analogue to the ``$\#$'' operator).  This creates a bijection between the spaces $\Gamma$ and $\mathcal{X}$, and they are therefore isomorphic. We are not aware of this construction made explicitly in the literature to represent the PCVF space $\mathcal{X}$; a similar construction is alluded to in~\cite{Poelke:2016}.

\paragraph{Packed and unpacked representations}

In order to naturally encode a null sum of $\gamma \in \Gamma$, in each face we only store the first two $\gamma$ values: $\gamma_{1,t}$ and $\gamma_{2,t}$. The choice is made without loss of generality---the choice of edges can be arbitrary, except that $e_2$ should follow $e_1$ in the counterclockwise order of the face. To reproduce all three when needed, we define an \emph{unpacking} operator $U$ as follows:
\begin{equation}
U_{|t}\begin{pmatrix} \gamma_{1,t} \\  \gamma_{2,t} \end{pmatrix} = \begin{pmatrix}  \gamma_{1,t} \\  \gamma_{2,t} \\ s_3\left(-s_{1,t}\gamma_{1,t}-s_{2,t}\gamma_{2,t}\right) \end{pmatrix}
\label{eq:packing-operator}
\end{equation}
The packed representation ``costs'' $2$ scalars per triangle, which is exactly the true dimension of $\mathcal{X}$. The effect of the packing operator $U^{-1}$ (in pseudo-inverse) is to simply throw away $\gamma_{3,t}$ if the null-sum condition is met: \mbox{$U_{|t}^{-1} \cdot U_{|t} = I_{2 \times 2}$}. We get that $U_{|t} \cdot U_{|t}^{-1}$ is a $3 \times 3$ matrix that filters away any non-null sum, while changing the $\gamma$ values if they violate it; we always avoid using it in this capacity.

With this representation, we reduce $P'$ to the operator we use in practice, $P$, where its pseudo-inverse $P^{-1}$ is an actual inverse, and both are defined as:
\begin{align}
P_{|t}&=\begin{pmatrix} e_{1,t} \\ e_{2,t} \end{pmatrix}  \nonumber \\
P^{-1}_{|t} &= \frac{s_1s_2}{2A_t}\begin{pmatrix} -e^{\perp}_{2,t} \\ e^{\perp}_{1,t} \end{pmatrix}^T.
\label{eq:packed-projection}
\end{align}
We use the convention $e_2^{\perp}=J_{|t}e_2$. $P$ and $P^{-1}$ aggregate the above per-face matrices into global operators. Note that $\left(e_1\right) \cdot \left(-e_2^{\perp}\right) = N_t \cdot \left(e_1 \times e_2\right)  = 2s_1s_2A_t$.  As such, we have $P \cdot P^{-1} = I_{2 \times 2}$ and $P^{-1} \cdot P$ is a $3 \times 3$ matrix that projects out the normal component from an ambient vector field in $\mathbb{R}^3$. We avoid the normal-component filtering capacity in our formulations here as well, and provide a proof that  $P^{-1} \cdot P$ is an identity for tangent vector fields in Appendix~\ref{app:pinv}.

%\paragraph{Null-sum projection} Note that if $P^{-1}$ is operated on edge values that do not obey the null-sum, it simply produces a vector in $\chi_h$ where the sum is filtered out. As such, we maintain that $P^{-1}P$ filters normal components, and consequently an identity in $\chi$, and that $P\cdot P^{-1}$ filters the average of and edge-based scalars to form a valid $\gamma \in \Gamma$, and consequently an identity in $\Gamma$.

\subsection{Halfedge Differential operators}

\begin{figure}
\includegraphics[width=0.4\textwidth]{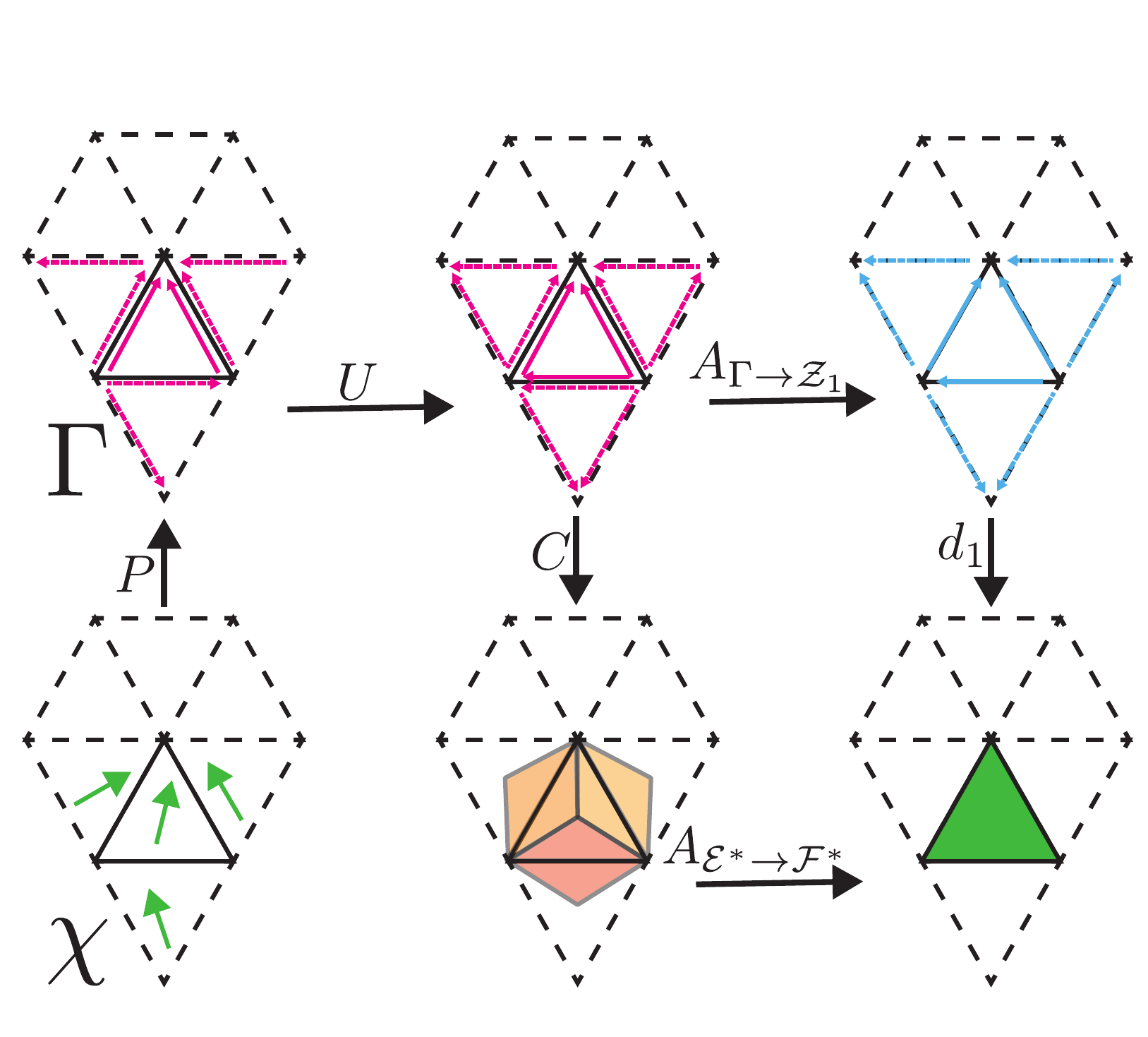}
\caption{The operators in the halfedge representation $\Gamma$.}
\label{fig:halfedge-calculus}
\end{figure}

We next redefine all differential operators for $\Gamma$ with the underlying paradigm that they should be equivalent to the operators in $\mathcal{X}$, albeit formulated in $\Gamma$ terms. We illustrate these operators in Figure~\ref{fig:halfedge-calculus}, and provide the entire set of differential operators for $\mathcal{X}$ in the $\Gamma$ setting in the rightmost columns of Table~\ref{table:all-operators}, comparing them with the analogous DEC and FEM operators.

\paragraph{Conforming gradient}

Consider the assignment operator $A_{\mathcal{Z}_1 \rightarrow \Gamma}$ that creates a halfedge form from a $1$-form by copying the associated oriented scalar on an edge to its two halfedges. We then get:
\begin{equation}
P^{-1} \cdot U^{-1} \cdot A_{\mathcal{Z}_1 \rightarrow \Gamma} \cdot d_0 f = G_\mathcal{V} f.
\end{equation}

The above relation demonstrates how DEC aligns with $\Gamma$ where exact $1$-forms, copied to halfedge forms, represent gradient fields---a fundamental parallel relation between DEC and $\mathcal{X}$. To avoid cumbersome notation, we denote $d_{0,\Gamma} = U^{-1} \cdot A_{\mathcal{Z} _1 \rightarrow \Gamma} \cdot d_0$, which is the differential operator of dimensions $2\left|\mathcal{F}\right| \times \left|\mathcal{V}\right|$ in $\Gamma$ space. 

We extend the DEC $d_1$ to be the (oriented) sum operator $d_1 = \sum_{i=1}^3{s_{i,t}\gamma_{i,t}}$ (working similarly to DEC $d_1$, except with the halfedges of the face rather than $1$-forms). To work with the packed form, we use $d_{1,\Gamma}=d_1\cdot U$. The null sum constraint is then encoded as the identity $d_{1,\Gamma}\cdot \gamma=0$ for every $\gamma \in \Gamma$.

The transpose operator $\left(A_{\mathcal{Z}_1 \rightarrow \Gamma}\right)^T \equiv A_{\Gamma \rightarrow \mathcal{Z}_1}$ creates $1$-forms from halfedge forms by summing up both halfedges scalars of each edge; we use it extensively in Section~\ref{subsec:mean-curl-representation}.

\paragraph{Curl}
We consider again $\gamma_{ik,1}$ and $\gamma_{ik,2}$, the two halfedge forms restricted to the edge $e_{ik}$ on the respective triangles $t_1$ and $t_2$. The curl operator $C: \Gamma \rightarrow \mathcal{E}^{\ast}$ is defined in $\Gamma$ space as: 
\begin{equation}
C_{|ik} =\gamma_{ik,1}-\gamma_{ik,2}.
\end{equation}
It is evident that curl-free fields in $\Gamma$ (or the equivalent $\mathcal{X}$) are such that the halfedge forms are equal on both sides of the edge, which means they are isomorphic to $1$-forms. As the null-sum constraint also dictates $d_{1,\Gamma}\gamma =0$ by definition, we have have that a curl-free $\gamma$ is isomorphic to a closed $1$-form. However, a halfedge form that is not curl free is not compatible with any DEC quantity. Since we represent $\Gamma$ with only two scalars per face, the complete definition for the curl operator is $C_\Gamma = C\cdot U$. Note that we have $C_\Gamma \cdot d_{0,\Gamma}=0$, which preserves the discrete structure of $\mathcal{X}$.

\subsection{Inner product}

The inner product between halfedge forms $\gamma_1, \gamma_2 \in \Gamma$ is  defined as:
\begin{equation}
\left(P^{-1}\gamma_1\right)^TM_{\mathcal{X}} \left(P^{-1}\gamma_2\right) =  \gamma_1^T \left( P^{-T} M_\mathcal{X} P^{-1}\right) \gamma_2 = \gamma_1^T M_\Gamma \gamma_2.
\label{eq:m-gamma-pmp}
\end{equation}

$M_\Gamma: 2\left|\mathcal{F} \right| \times 2\left|\mathcal{F} \right|$ has the following simple structure:

\begin{equation}
M_{\Gamma | t} =  \frac{1}{2}\left(U_{|t}\right) ^T\begin{pmatrix} cot(\alpha_1) & & \\ & cot(\alpha_2) & \\  & & cot(\alpha_3) \end{pmatrix} U_{|t},
\label{eq:m-gamma}
\end{equation}

where $\alpha_j$ is the angle opposite edge $j$ in face $t$, and $U$ is the unpacking operator as before. Simply put, we get a diagonal mass matrix for the unpacked null-summed $\gamma \in \Gamma$. We show the proof in Appendix~\ref{app:m-gamma}. Equipped with these basic operators, the divergence and Laplacian can be directly defined as in Table~\ref{table:all-operators}.

%\paragraph{Divergence and Laplacian} Equipped with the inner product, the (conforming) divergence is defined as:
%\begin{equation}
%D_\Gamma = \left(d_{0,\Gamma}\right)^T  M_\Gamma.
%\end{equation}

%Remember that $M_\Gamma$ already contains the unpacking operator $U$, so the sizes of the matrices in the product are compatible with each other. The (integrated) Laplacian is then:

%\begin{equation}
%L_\Gamma:\mathcal{V} \rightarrow \mathcal{V^{\ast}} := \left(d_{0,\Gamma}\right)^T %\cdot M_\Gamma  \cdot d_{0,\Gamma},
%\end{equation}

%which is exactly the cotangent Laplacian, as expected, since $\Gamma$ represents $\mathcal{X}$.

%\paragraph{Rotated cogradient} It is straightforward to show that the expression $JG_{\mathcal{E}}$ in $\mathcal{X}$ is equivalent to $-M_\Gamma^{-1} C_\Gamma^T$. We then have:
%$$-D_\Gamma \cdot M_\Gamma^{-1} C_\Gamma^T = -d_{0,\Gamma}^T \cdot C_\Gamma^T = %0,$$ 
%which means we preserve the \changed{structure of the dual} sequence as well.

%However, some of them require clarification, which we bring in the following.

\subsection{Mean-curl representation}
\label{subsec:mean-curl-representation}

Though the halfedge forms $\gamma \in \Gamma$ are equivalent to PCVFs in $\mathcal{X}$ through the projection operator $P$, we need  an alternative and equivalent representation for them that reveals their differential properties, to be used in our subdivision schemes. Given the two halfedge forms $\gamma_{ik,1}$ and $\gamma_{ik,2}$ on both sides of edge $ik$ adjacent to triangles $t_1$ and $t_2$ in our usual notation, we define:
\begin{align}
z_{1|ik}&=\frac{\gamma_{ik,1}+\gamma_{ik,2}}{2} \Rightarrow z_1 = \frac{1}{2}A_{\Gamma \rightarrow \mathcal{Z}_1}\cdot U \cdot \gamma \\ \nonumber
\epsilon_{|ik}&=\frac{\gamma_{ik,1}-\gamma_{ik,2}}{2} \Rightarrow \epsilon = \frac{1}{2}C_\Gamma \cdot \gamma
\end{align}

In words, $z_{1}$ is the DEC $1$-form that is the mean of the two halfedge forms, and $\epsilon$ is half of the FEM curl of $\gamma$. This representation is trivially equivalent to that of the unpacked $\gamma$. We denote the conversion operator as $W$ as follows:
\begin{align}
\begin{pmatrix} z_1 \\ \epsilon \end{pmatrix} &= W \cdot \gamma = \frac{1}{2}\begin{pmatrix} A_{\Gamma \rightarrow \mathcal{Z}_1} \\ C \end{pmatrix} U \cdot \gamma,\nonumber \\ 
W^{-1} &= U^{-1} \begin{pmatrix} A_{\Gamma \rightarrow \mathcal{Z}_1}^T & C^T \end{pmatrix}.
\label{eq:W}
\end{align}
 %\BAC{I think the minus before $C^T$ should not be there}
Note that $z_1 \in \mathcal{Z}_1$ is a signed oriented quantity while $\epsilon \in \mathcal{E^{\ast}}$ is an unsigned integrated quantity.  We emphasize that the null-sum constraint $d_{1,\Gamma} \cdot \gamma =0$ does not imply that $z_1$ is curl-free in the DEC sense. That is, we do not have $d_1z_1=0$ in general.%; the only exception is when the equivalent vector field in $\mathcal{X}$ is FEM curl-free, where $\epsilon = 0$. \BAC{this statement is a repetition of the statement at the end of the curl section? Also, the inverse is not exactly that when considering triangulations with boundary.}

\paragraph{Null sum constraint in mean-curl} The mean-curl representation is not trivially equivalent to the packed $\Gamma$ we use, since it has values for all edges, whereas $\gamma$ is spanned by two halfedges within each triangle (hence the use of $U^{-1}$ in $W^{-1}$). To get equivalence, we need to formulate the $\Gamma$ null-sum requirement with $\left(z_1,\epsilon\right)$. This formulation has a surprisingly elegant form. Consider the face $t=ijk$, and the signs $s$ for the coincident halfedge forms $\gamma$. Then: 
\begin{align}
d_{1,\Gamma|t}\cdot \gamma_{|t}=0=s_{ij}\gamma_{ij}+s_{jk}\gamma_{jk}+s_{ki}\gamma_{ki} &= \\ \nonumber
s_{ij}z_{ij}+s_{jk}z_{jk}+s_{ki}z_{ki} - \epsilon_{ij} -  \epsilon_{jk} - \epsilon_{ki} &= \\ \nonumber
d_{1|t}z_{1|t} - A_{\mathcal{E^{\ast}} \rightarrow \mathcal{F}|t}\epsilon_{|t}
\end{align}

where $d_{1|t}$ is the DEC $d_1$ operator restricted to $f$, and $A_{\mathcal{E^{\ast}} \rightarrow \mathcal{F^{\ast}}}$ is the summation operator $\epsilon_{ij}+\epsilon_{jk}+\epsilon_{ki}$ (analogous to $A_{\Gamma \rightarrow \mathcal{Z}_1}$). In global notation the null-sum constraint reads:
\begin{equation}
\label{eq:mean-curl-null-sum}
d_1z_1 - A_{\mathcal{E^{\ast}} \rightarrow \mathcal{F}^{\ast}}\cdot \epsilon = 0.
\end{equation}
Note again that when $\epsilon$ is $0$, $z_1$ is a closed $1$-form and we get the DEC identity $d_1z_1 = 0$. More generally, as the DEC definition of curl (see Table~\ref{table:all-operators}) is exactly $d_1z_1$, the DEC face-based curl of the mean $1$-form $z_1$ is then nothing but the face-summed edge-based FEM curl of the underlying field $\gamma$. We are not aware of this connection between DEC curl and FEM curl pointed out before.

%The DEC divergence of the mean $1$-form $z_1$ is tied to $\epsilon$ and $\gamma$ as follows:

%\begin{align}
%d_0^TM_1^{-1}z_1 = 
%end{align}

The mean-curl representation reveals important ties between DEC and FEM more clearly:

\begin{itemize}
\item $\gamma$ is FEM-exact if and only if  $z_1$ is DEC-exact with the same function $f \in \mathcal{V}$ so that $d_0f = z_1$, and where $\epsilon=0$.
\item $\gamma$ is FEM-harmonic if and only if $z_1$ is DEC-harmonic. This is straightforward to see, as the DEC divergence operators $d_0^TM_1$ and $D_\Gamma$ identify when $\epsilon = 0$.
\item FEM-coexact $\gamma$ does not correspond to coexact $z_1$; this is evident by the incompatible dimensions of the spaces. However, suppose that $\epsilon \in \mathcal{E}^{\ast}$ is the curl of $\gamma$, then we have in this case a simple expression for the divergence of $z_1$: $d_0^T M_1 z_1= D_{\Gamma} C^T \epsilon$.
\end{itemize}

%To summarize, harmonic and exact spaces are equivalent in FEM and DEC. Coexact spaces are not the same, since they have different dimensions. However, the DEC-curl of the averaged $1$-form of some FEM field is just the average of the FEM-curl over faces.

\paragraph{Discussion: refinable Hodge decomposition}

Given the insights of the mean-curl representation, there is a subtle, yet important, distinction between the way DEC and FEM treat the Hodge decomposition, which we need to make in order to properly define subdivision for PCVFs in $\mathcal{X}$. The DEC Hodge decomposition factors a $1$-form $z_1 \in \mathcal{Z}_1$ into pointwise $z_0 \in \mathcal{V}$, harmonic part $z_h$, and \emph{integrated} $z_2 \in \mathcal{F}^{\ast}$ (the equivalent of $\mathcal{Z}_2$). They further rely on refinable function spaces to perform subdivision (Section~\ref{subsec:subdivision-exterior-calculus}). For this, using integrated $\mathcal{F}^{\ast}$ is the correct choice, since $\mathcal{Z}_2$ admits a natural refinable hierarchy by triangle quadrisection. The pointwise dual $2$-forms do not admit a refinable structure in this manner, and subdividing them directly would constitute as a ``variational crime''.

However, the FEM Hodge decomposition classically uses the pointwise elements in $\mathcal{E}$ to span its coexact part, which is, similarly to the dual 2-form space $\mathcal{Z}^{\ast}_2$, not a refinable space. Nevertheless, the Hodge decomposition can be  defined in $\Gamma$ analogously to DEC by using $f \in \mathcal{V}$, (half) curl $\epsilon \in \mathcal{E^{\ast}} $, and harmonic $h \in H_\Gamma$ as follows:
$$
\forall \gamma \in \Gamma,\ \exists f \in \mathcal{V},\ \epsilon \in \mathcal{E}^{\ast},\ h \in H_\Gamma:\ \gamma = d_{0,\Gamma} f + 2 M_\Gamma^{-1}C^T L_\mathcal{E}^{-1} \epsilon + h.
$$

Other than just for revealing algebraic relations between FEM and DEC, we use the halfedge representation, mostly in its mean-curl representation, to establish PCVF subdivision schemes.

% !TEX root =  SubdivisionDirectionalFields.tex

\begin{figure}[h!]
	\centering
	\includegraphics[scale=1]{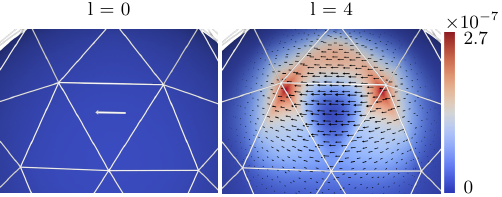}
	
	\caption{A refined ($l=4$) Basis function from a single coarse ($l=0$) unit-length vector. The color-coding on the fine level depicts the per-face Hodge energy component $\left|C_\Gamma \gamma \right|^2 + \left|D_\Gamma \gamma \right|^2$ of the field, averaged to the vertices for visualization purposes. The glyph arrows on the fine level visualize directions and relative magnitudes.}%\BAC{Color $l=0$ as well?}}
	\label{fig:basis-function}
\end{figure}

\section{Subdivision Vector Fields}
\label{sec:subdivision}

Our purpose in constructing subdivision schemes for halfedge forms is the ability to work with PCVFs in a multi-resolution structure-preserving manner. Specifically, we work with subdivided vector fields on very fine subdivided meshes, that are restricted to low-dimensional fields defined with the coarse control cage, for purposes of efficiency and robustness.
We define $S_\mathcal{V}$ as the Loop subdivision matrix for vertex-based quantities, and $S_\mathcal{F^{\ast}}=S_2$ as the half box spline face-based subdivision matrix, equivalent to $S_0$ and $S_2$ in the SEC scheme. 
For halfedge-based subdivision, we construct three distinct and interrelated operators for each subdivision level $l$:

\begin{itemize}
\item $S^l_1$ for $1$-forms, of dimensions $\left| V^{l+1} \right| \times \left|  V^l\right|$.
\item $S^l_\mathcal{E^{\ast}}$, for unsigned integrated edge-based quantities (like curl), of dimensions $\left| E^{l+1} \right| \times \left| E^l \right|$.
\item $S^l_\Gamma$, for halfedge forms composed of both. It is then of dimensions $2\left| F^{l+1} \right| \times 2 \left|F^l \right|$.
\end{itemize}

$S_1$ is defined as in SEC (except our boundary modifications; see auxiliary material), so we need to define the latter two. For clarity, we often omit the level indicator $l$, as the operators are stationary, and the level can be understood from the context. We provide the full set of stencils in the auxiliary material.

In order to define structure-preserving operators on $\Gamma$, we require that $S_\Gamma$ and $S_{\mathcal{E}^{\ast}}$ obey the following  commutation rules:
\begin{align}
\label{eq:commutation}
d_{0,\Gamma} \cdot S_\mathcal{V} &= S_{\Gamma} \cdot  d_{0,\Gamma} \\ \nonumber
C_\Gamma \cdot S_\Gamma  &=  S_\mathcal{E^{\ast}} \cdot C_\Gamma.
\end{align}
In words, subdivided halfedge forms that represent gradient fields should result in gradient fields of the subdivided vertex-based scalar function, and the curl of a subdivided vector field should be equal to the subdivided curl of a vector field. 
To satisfy these conditions, our subdivision matrix for halfedge-forms is defined directly on the mean-curl representation as follows:
\begin{equation}
S_{\Gamma} \cdot \gamma = W^{-1} \begin{pmatrix} S_1 & 0 \\ 0 & S_\mathcal{E^{\ast}} \end{pmatrix}  \begin{pmatrix} z_1 \\ \epsilon \end{pmatrix} = W^{-1} \begin{pmatrix} S_1 & 0 \\ 0 & S_\mathcal{E^{\ast}} \end{pmatrix} W \cdot \gamma,
\end{equation}
Since $S_\Gamma$ is defined with the mean-curl representation which is in unpacked form, we need to verify that the null-sum requirement for the subdivided field $\left(z_1^{l+1},\epsilon^{l+1}\right)$ is satisfied before the application of $W^{-1}$, or otherwise $W^{-1}$ will project the result unto the null-summed space $\Gamma$ and the requirements in Equation~\ref{eq:commutation} will not result in the promised structure-preserving properties. That is, we require (as per Equation~\ref{eq:mean-curl-null-sum}):
$$
d_1z_1^{l+1} - A_{\mathcal{E}^{\ast} \rightarrow \mathcal{F^{\ast}}}\epsilon^{l+1}=0.
$$
As we inherit (albeit with some slight modifications) $S_1$ and $S_{\mathcal{F}^{\ast}}$ from SEC, our degrees of freedom for the requirements are in the definition of $S_\mathcal{E^{\ast}}$. To satisfy all requirements, we design it to adhere to the following additional commutation relation:
\begin{equation}
S_\mathcal{F^{\ast}}A_{\mathcal{E^{\ast}} \rightarrow \mathcal{F^{\ast}}} = A_{\mathcal{E^{\ast}} \rightarrow \mathcal{F^{\ast}}}S_\mathcal{E^{\ast}}.
\end{equation}
In words, the face-based average of the subdivided curl should be equal to the subdivided face-based average of the coarse curl. This commutation elegantly preserves the null-sum requirement, as for level $l$, with mean $z_1^l$ and half-curl $\epsilon^l$, we get
\begin{align} \nonumber
\textit{(Level $l$ null-sum\ constraint (Eq.~\ref{eq:mean-curl-null-sum}))}&\ \ \ d_1z_1^l - A_{\mathcal{\mathcal{E}^{\ast}} \rightarrow \mathcal{F}^{\ast}} \epsilon^l =0 \Rightarrow \\ \nonumber
\textit{(Subdivision)}&\ \ \  S_\mathcal{F^{\ast}}d_1z_1^l - S_\mathcal{F^{\ast}} A_{\mathcal{\mathcal{E}^{\ast}} \rightarrow \mathcal{F}^{\ast}} \epsilon^l =0 \Rightarrow \\ \nonumber  
\textit{(Commutation)}&\ \ \  d_1S_1z_1^l - A_{\mathcal{\mathcal{E}^{\ast}} \rightarrow \mathcal{F}^{\ast}} S_\mathcal{\mathcal{E}^{\ast}}\epsilon^l =0  \Rightarrow \\
\textit{(Level $l+1$ null-sum\ constraint)}&\ \ \  d_1z_1^{l+1} - A_{\mathcal{E}^{\ast} \rightarrow \mathcal{F^{\ast}}}\epsilon^{l+1}=0.
\end{align}

%\item Due to the linear commutation relation, we conjecture that $S_\mathcal{E^{\ast}}$ converges to a smooth curl in the same order of smoothness and of convergence as $S_\mathcal{F^{\ast}}$.
%\end{enumerate}

\begin{figure}[h!]
	\centering
	\includegraphics[scale=1]{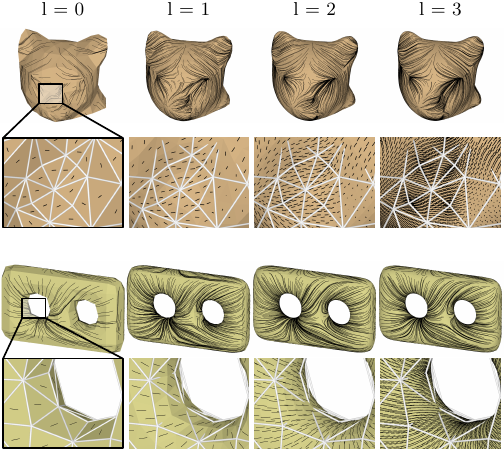}
	\caption{Multiple levels of subdivision vector fields on the cathead (top, genus  0) and bitorus (bottom, genus  2) models. Note that the subdivsion preserves the features (sources, sinks, and vortices) of the fields.}
	\label{fig:subdivision-vector-fields}
\end{figure}

Having secured the null-sum constraint, we can safely use $W^{-1}$ to get the fine level field $\gamma^{l+1}$, where all the promised differential properties are guaranteed.
We show an example of a basis function of the subdivision operator in Figure~\ref{fig:basis-function}, and some examples of full subdivision vector fields in Figure~\ref{fig:subdivision-vector-fields}.

\subsection{Boundary behavior}

\begin{figure}[h!]
	\centering
	\includegraphics[width=0.5\textwidth]{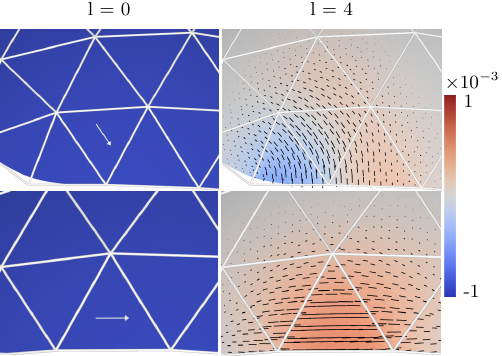}
	\caption{Basis functions for coarse and fine levels near the boundary. The initial vector is of unit length. The color-coding depicts the local face averaged curl $A_{\mathcal{E^{\ast}} \rightarrow \mathcal{F^{\ast}}}C\cdot\gamma$.}
	\label{fig:basis-boundary}
\end{figure}

Our concepts of halfedges and the differential operators do not trivially extend to meshes with boundaries. Recall that our reasoning for subdivision is to commute with the gradient and the curl operators. However, the discrete curl operator on the boundary is not well-defined for a single edge: consider a boundary face $t=ijk$ with boundary edge $e_{ij}$, and the associated halfedge form $\gamma_{|ij}$. As studied in~\cite{Poelke:2016}, the Hodge decomposition for meshes with boundaries admits several valid choices for decomposition, culminating in either Dirichlet or Neumann boundary conditions. We choose to assume that a function $f\in \mathcal{V}$ is defined everywhere, including the boundary, and that we commute with its gradient. Consequently, we assume that the boundary curl is zero by definition. That is, on the boundary, we define $z_{1|ij}=\gamma_{ij}$ and $\epsilon_{|ij}=0$. We adapt $W$ and $W^{-1}$ accordingly, noting that $z_{1|ij}$ is the only contribution to the field for boundary edge $ij$. Our subdivision matrices are designed to reflect that, where $S_\mathcal{E^{\ast}}$ reproduces zero curl on the boundary, and $S_1$ is redefined to preserve the null-sum with this constrained $S_\mathcal{E^{\ast}}$. We show an illustration of boundary vector field basis functions on the boundary in Figure~\ref{fig:basis-boundary}.

\subsection{SHM differential operators}
Following the reasoning of Section~\ref{subsec:subdivision-exterior-calculus}, we restrict $M_\Gamma$ from a fine mesh back to a coarse mesh as follows:
\begin{equation}
\mathbb{M}^0_{\Gamma} = \left(\mathbb{S}^l_\Gamma\right)^T \cdot M^l_{\Gamma} \cdot \mathbb{S}^l_\Gamma.
\end{equation}
By this process of mass-matrix restriction, we process fine-level PCVFs that are spanned by the low-dimensional subdivided coarse-level PCVFs, directly on the coarse mesh. In analogy to SEC, we denote this technique as \emph{Subdivision Halfedge-form Method} (\textbf{SHM}).

By the commutation relations, the subdivided SHM curl of a field is equal to the fine curl, and when a field is SHM-exact on the coarse mesh, then it is also FEM-exact on the fine mesh, where the fine function is the subdivision of the coarse one. Nevertheless, the SHM divergence behaves differently from the fine FEM divergence, as:
\begin{align}
 \nonumber
\mathbb{D}_\Gamma^0\gamma^0 = d_{0,\Gamma}^T \mathbb{M}^0_{\Gamma} \gamma^0 &= \\  \nonumber
d_{0,\Gamma}^T \left(\mathbb{S}^l_\Gamma\right)^T \cdot M^l_{\Gamma} \cdot \mathbb{S}^l_\Gamma\gamma^0 &= \\ 
\left(\mathbb{S}^l_\mathcal{V}\right)^Td_{0,\Gamma}^T \cdot M^l_{\Gamma} \gamma^l &=\left(\mathbb{S}^l_\mathcal{V}\right)^TD^l\gamma^l.
\label{eq:divergence-bleeding}
\end{align}

\begin{figure}[h!]
	\centering
	\includegraphics[scale=1]{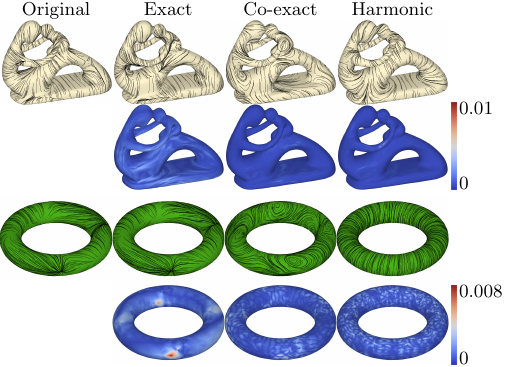}
	\caption{SHM Hodge decomposition on models with genus 6 (first row) and 1 (third row), where a field is decomposed with SHM operators, subdivided, and shown in streamlines. The second and fourth row show the absolute value of the fine-level divergence, from which the high-frequency divergence pollution in the co-exact and harmonic part is evident. The fields are given by $\vec{v}_t(x,y,z) = \lbrace\sin(\pi x)y, \sin(\pi x y)/r^2,\cos(\pi z) + x^2 + y^2\rbrace$, where $x,y,z$ are the coordinates of the face barycenter and $r^2=x^2+y^2+z^2$. We add a random harmonic field from the null space of the Hodge Laplacian to the original field, and it is reproduced in the decomposition.}
	
	%To the torus, the two harmonics are added with scales 0.6 and 0.03, which are acquired via QR decomposition. To the top model, the harmonics were added with scales 0.0632, 0.7607, 0.5923, 0.8572, 0.9971, 0.7686, 0.2238 and 0.0579.}
\label{fig:hodge-decomposition}
\end{figure}

Note that we use $\mathbb{D}_\Gamma$ to denote the SHM divergence operator in line with other notation. In words, the divergence of a subdivided field is equal to the divergence of the resulting coarse field only through the restriction $\left(\mathbb{S}^l_\mathcal{V}\right)^T$. That essentially means that the divergence of the fine field might have ``high frequency'' components in $\kerM\left(\mathbb{S}^l_\mathcal{V}\right)^T$ (see Figure~\ref{fig:hodge-decomposition}). This is an analogous phenomenon to the divergence pollution of SEC (Section~\ref{subsec:subdivision-exterior-calculus}). Note that the structure of SHM is preserved notwithstanding: SHM-exact fields are SHM curl-free, and SHM-coexact fields are SHM-divergence free.

The restricted mass matrices $\mathbb{M}$ are not diagonal anymore due to the two-ring support of any $\mathbb{S}$. Additionally, some operators are defined with inverse mass matrices, which are dense and non-local. In practice, we almost never need to compute the exact inverse, and we show how to circumvent this problem in the relevant applications.

\paragraph{Hodge decomposition}

In Figure~\ref{fig:hodge-decomposition} we show a Hodge decomposition of a procedurally-generated field with the SHM operators, subdivided to a fine level $l=3$. It is evident that the exact part subdivides as defined, but also that there is high-frequency divergence that pollutes the co-exact and the harmonic parts.

\begin{figure}
	\centering 
	\includegraphics[scale=1]{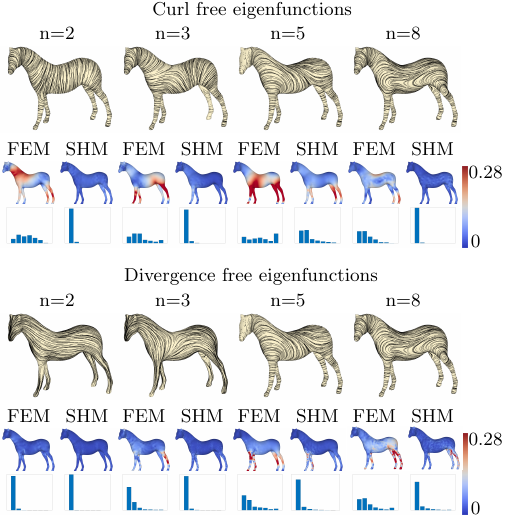}
	
	\caption{Top: exact, and bottom: co-exact eigenfunctions of the SHM ($l=3$) Hodge Laplacian, subdivided and visualized at the fine level. The color-coding denotes the norm difference $||P^{-1}\gamma'_t - P^{-1}\gamma_t^l||_2$ per triangle $t$, where $\gamma'_t$ is the eigenfunction subdivided from the coarse level (FEM and SHM) eigenfunction to the fine level (with normalization so that $\gamma_t^{'T} M^l_\Gamma \gamma'_t=1$), and $\gamma_t^l$ is the ground-truth eigenfunction of the fine level Hodge Laplacian. The color scale depicts this pointwise error.}  %For curl-free eigenfunctions, we normalize the fine level by dividing by its $L_2$ product and then taken to $\Gamma$. For $\psi$, we first calculate the curl of the field corresponding to the coarse level eigenfunction, take this to the fine level, normalize by dividing by the $L_2$ product and then compute the corresponding $\gamma$ field.}}
	\label{fig:hodge-eigs}
\end{figure}

\begin{figure}
	\centering
	\includegraphics[scale=0.8]{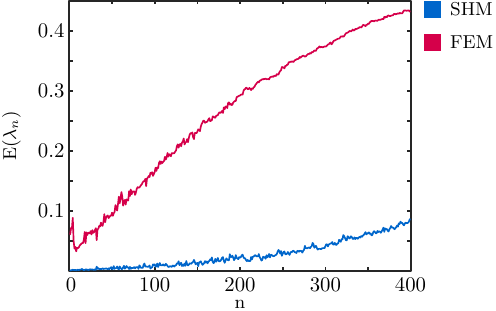}
	
	\caption{Comparing eigenvalues between the fine-level conforming part of the Hodge Laplacian and the SHM or coarse-FEM eigenvalues, where the error is calculated for SHM as $E(\lambda_n) = |\lambda_{n,SHM} - \lambda_{n,\text{fine}}|/|\lambda_{\text{n,fine}}|$, and similarly for FEM. The same mesh as Fig.~\ref{fig:hodge-eigs} is used, with $|V| = 752$ at $l=0$ and $|V|=48002$ at $l=3$. We note that a concrete proof for this relation appears in~\cite{Shoham:2019}.}
	\label{fig:hodge-L-eigs-diff}
\end{figure}

\begin{figure}
	\centering
	\includegraphics[scale=0.8]{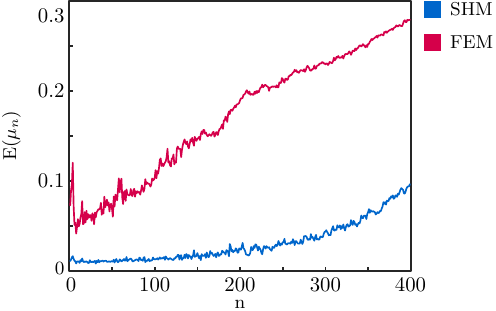}
	
	\caption{Similar comparison as in Figure ~\ref{fig:hodge-L-eigs-diff} for the non-conforming part of the Hodge Laplacian, where the error in SHM is $E(\mu_n) = |\mu_{n,SHM} - \mu_{n,\text{fine}}|/|\mu_{\text{n,fine}}|$, and similarly for coarse FEM.  The same mesh has $|E|=2250$ at $l=0$ and $|E|=144000$ at $l=3$.}
	\label{fig:hodge-CL-eigs-diff}
\end{figure}

\paragraph{Hodge Spectrum}
The spectrum of the PCVF Hodge Laplacian $L_\mathcal{X}$ is studied in~\cite{Brandt:2016}, where they show that the spectrum of $L_\mathcal{X}$ comprises harmonic fields (in its null space), gradients of eigenfunctions of $L_\mathcal{V}$, and cogradients of eigenfunctions of $L_\mathcal{E}$. Using the SHM mass matrices, these relations still hold for the SHM Hodge Laplacian $\mathbb{L}_\Gamma$:
\begin{align}
\nonumber
\forall \phi \in \mathcal{V},~\lambda \in \mathbb{R},\ s.t.\ \mathbb{L}_\mathcal{V}\phi &= \lambda \mathbb{M}_\mathcal{V} \phi \Rightarrow \mathbb{L}_\Gamma \cdot d_{0,\Gamma} \cdot \phi = \lambda \mathbb{M}_\Gamma \cdot d_{0,\Gamma} \cdot \phi. \\
\forall \psi \in \mathcal{E},~\mu \in \mathbb{R},\ s.t.\ \mathbb{L}_\mathcal{E}\psi &= \mu \mathbb{M}_\mathcal{E} \psi \Rightarrow \mathbb{L}_\Gamma \cdot \mathbb{M}_\Gamma^{-1} C_\Gamma^T \cdot \psi = \mu C_\Gamma^T \cdot \psi.
\end{align}
Note that a term of $\mathbb{M}_\Gamma \cdot \mathbb{M}_\Gamma^{-1}$ was simplified in the right-hand side of the last equation. We used subdivision level $l=3$, and computed the SHM Hodge eigenfunctions for several eigenvalues. We compare them against the ground-truth fine eigenfunctions in Figure~\ref{fig:hodge-eigs}. In Figures~\ref{fig:hodge-L-eigs-diff} and~\ref{fig:hodge-CL-eigs-diff}, we further analyze the relative error between the fine spectrum and the FEM and SHM spectra for the Hodge Laplacian. As can be seen, the SHM spectrum is a much better approximation of the fine Hodge spectrum than the coarse FEM one, for more than half of the full spectrum.

\paragraph{Errors and convergence}
To study the behavior of our PCVF subdivision, we look at the behaviour of the SHM Hodge Laplacian for the vector equation:
$$
\mathbb{L}_\Gamma \cdot \gamma = b,
$$
where $b \in \Gamma$ is some given field and $\mathbb{L}_\Gamma$ is the SHM Hodge Laplacian. We conduct two error and convergence tests as follows.

\begin{figure*}[h!]
	\centering
	\includegraphics[width=\textwidth]{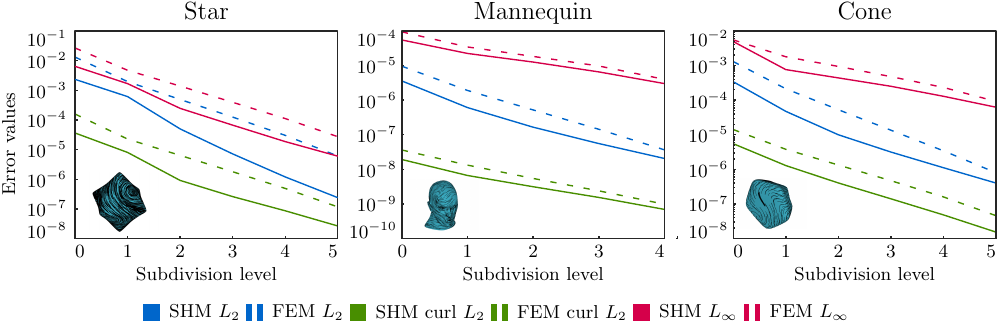}
	\caption{$L_2$, $L_\infty$, and the $L_2$ error of the curl, for the projection error of the vector Poisson equation on three models, as a function of subdivision level. We measure $L_\infty = \max |\delta \gamma|$ and $L_2 = \sqrt{\nicefrac{(\delta \gamma)^T M_{\Gamma} \delta\gamma}{\sum M_\Gamma}}$, where $\delta\gamma = \bar{\gamma}^l-\gamma^l$, with $\bar{\gamma}^l$ as the ground-truth solution and $\gamma^l$ as the subdivided solution. The curl $L_2$ error is measured as $\sqrt{\nicefrac{(C \delta \gamma)^T M_{\mathcal{E}^\ast}C \delta\gamma}{\sum M_{\mathcal{E}^\ast}}}$. We use $l=6$ for the Cone and Star models, and $l=5$ for the Mannequin.
	}
	\label{fig:projection-error}
\end{figure*}

\textbf{Projection error:} we measure the error that is obtained by approximating the fine-level FEM with the low-dimensional SHM. For this, we choose the right-hand $b^0$ procedurally on some coarse mesh (level 0), and subdivide it several times to get $b^l$, where we consider the solution $\overline{\gamma}^l$ to the Hodge Laplacian system with this right-hand as the ground-truth reference. For each level $0 \leq k < l$ we solve for $\mathbb{L}^{k}_\Gamma \gamma^k = b^k$, where $\mathbb{L}^{k}_\Gamma$ is the SHM Hodge Laplacian at level $k$ restricted from level $l$. We then subdivide $\gamma^k$ to get $\gamma^{k\rightarrow l}$,  and measure the $L_2$ and $L_\infty$ error against the ground-truth solution $\overline{\gamma}^l$. For reference, we compare to a regular FEM solution at level $k$, computed as $L^{k}_\Gamma \gamma'^k = b^k$, also subdivided to level $l$ and measured against the ground-truth solution. We show the results in Figure~\ref{fig:projection-error}, and analyze convergence rates in Table~\ref{table:proj-err-slopes}. It is evident that the SHM solution has superior performance in terms of error when compared against the regular FEM solution, almost consistently with 1--2 order of magnitudes less error. Interestingly enough, the convergence rates are similar.

 %We test the projection error of our SEM operators: the error of using a low-dimensional  of the fine function space. For each level  we compute $\gamma$ for the SEM Laplacian $\mathbb{L}^k_\Gamma$ and the FEM Laplacian $L^k_\Gamma$. Then, we subdivide $\gamma^k$ to $\gamma^l$.  for two different meshes. As can be seen, the SEM error is lower than the FEM error and converges to a small error. 
\textbf{Operator error:} we measure the error that is obtained on the coarse level $l=0$ operator, by restricting the SHM operators only from a level $k < l$, rather than from the fine level $l$ on which we wish to work. For instance, regular FEM operators are used when $k=0$ and the full SHM when $k=l$. We show the result in Table~\ref{table:operator-error}. As evident, the operator error diminishes quickly in the very coarse levels, but then it plateaus to a reasonable error. This suggests that a good approximation for processing on level $l$ can be accomplished with a fairly low SHM level $k$; that can be explained by the rapid convergence of subdivision schemes~\cite{Dahmen:1986}. 
\begin{table}[ht!]
\begin{tabular}{|c|c|c|c|c|c|c|}
\hline
Level & \multicolumn{2}{|c|}{Cone} & \multicolumn{2}{|c|}{Mannequin}  & \multicolumn{2}{|c|}{Star} \\
\hline
 & $L_\infty$ & $L_2$ & $L_\infty$ & $L_2$ &$L_\infty$ & $L_2$ \\
\hline
0 & 10.8 & 7.22 & 0.229e-1 & 0.460e-4 & 2.02 & 0.618 \\ 
1 & 4.88 & 0.326 & 0.145e-1 & 0.724e-5 & 0.578 & 0.496e-1 \\ 
2 & 4.48 & 0.221 & 0.148e-1 & 0.631e-5 & 0.492 & 0.396e-1 \\ 
3 & 4.46 & 0.222 & 0.148e-1 & 0.638e-5 & 0.468 & 0.409e-1 \\ 
4 & 4.47 & 0.225 & 0.149e-1 & 0.644e-5 & 0.461 & 0.417e-1 \\ 
5 & 4.47 & 0.227 & - & - & 0.459 & 0.420e-1 \\ 
\hline
\end{tabular}
\caption{Operator $L_2$ and $L_\infty$ errors for the three models. } 
\label{table:operator-error}
\end{table}

\begin{table}[h!]
\centering
\begin{tabular}{|c|c|c|c|}
\hline
\textbf{Error} & \multicolumn{3}{c|}{\textbf{Model}}\\
\hline
&Star & Mannequin & Cone\\
\hline
SHM $L_2$ & 2.54 & 1.77 & 1.82 \\ 
FEM $L_2$ & 2.00 & 1.91 & 2.00 \\ 
SHM $L_\infty$ & 1.90 & 0.978 & 1.10 \\ 
FEM $L_\infty$ & 1.80 & 1.06 & 1.07 \\ 
SHM Curl $L_2$ & 1.95 & 1.13 & 1.59 \\ 
FEM Curl $L_2$ & 1.87 & 1.24 & 1.54 \\ 

\hline
\end{tabular}
\caption{Projection error convergence rates for the models in Fig. \ref{fig:projection-error}. The values correspond to the convergence factor $\beta$ for the error hypothesis $a_0h^{-\beta}$, where $h$ is the mean edge length of the mesh.Fine levels $l$ are at $6$ for the Cone and Star experiment and $5$ for the Mannequin experiment.}
\label{table:proj-err-slopes}
\end{table}

% !TEX root =  SubdivisionDirectionalFields.tex

\section{Subdivision \texorpdfstring{$N$}{N}-directional fields}
\label{sec:directionals}

We next extend our subdivision operators to $N$-directional subdivision, with the same structure-preserving guarantees. We do so by applying the local reduction of such fields into single-vector fields on branched cover spaces, which are introduced in~\cite{Lawrence:2018}.

\begin{figure}
	\centering
	\includegraphics[width = 0.45\textwidth]{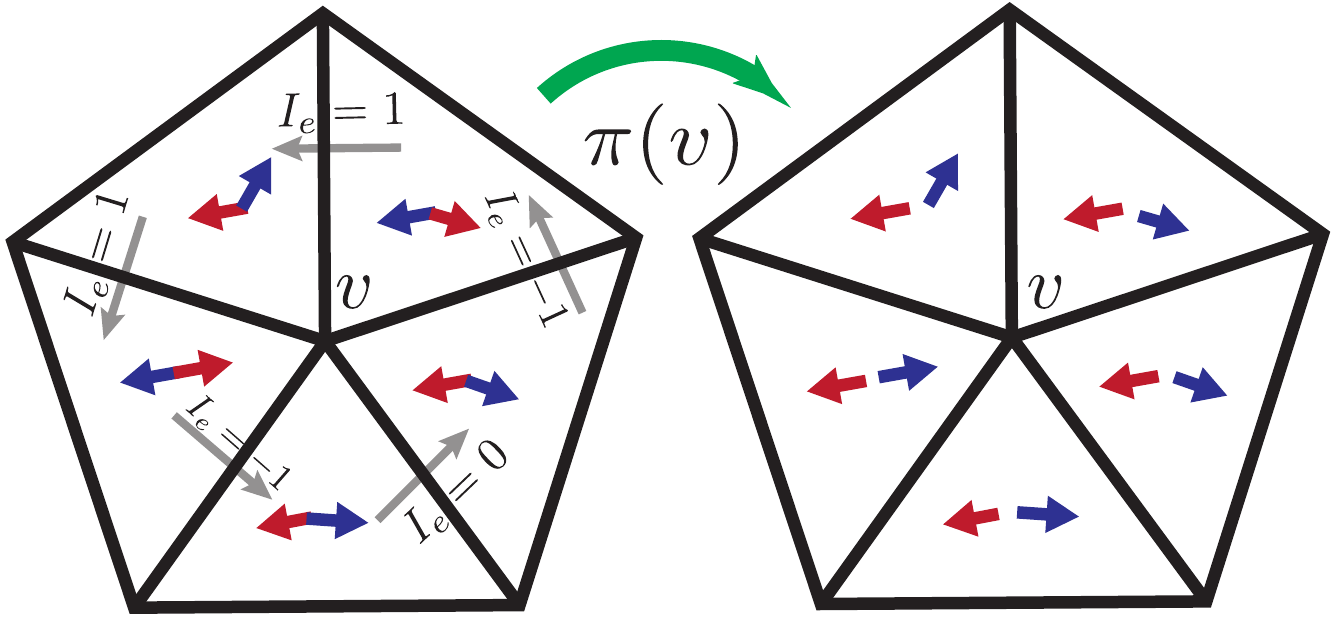}
	\caption{Matching and combing. Left: a non-singular matched $1$-ring. Right: as the $1$-ring is non-singular, applying $\Pi(v)$ results in a combed separated field with a trivial matching.}
	\label{fig:matching-combing}
\end{figure}

We work with $N$-directional fields that are elements of $\mathcal{X}^N$: in every face $t$ there are $N$ indexed vectors $\left\{v_{t,1},\ldots, v_{t,N}\right\}$, not necessarily symmetrically ordered. We assume that the field is equipped with a \emph{matching}: a map between the vectors on a face $t_1$ to those in an adjacent face $t_2$, associated with the dual edge $e$ between them.  Furthermore, we assume the matching is (index) \emph{order-preserving}: the matching is parameterized by a per-edge index $I_e$, where a vector of index $k$ on face $t_1$ is matched to vector of index $k+I_e$ (modulo $N$) on face $t_2$ (see Figure~\ref{fig:matching-combing}). We denote the full matching as $I_E$.

The indices of the vertices are defined as $I_V=\frac{1}{N}d_0^T\cdot I_e$~\cite{Crane:2010}, as $d_0^T$ is the DEC boundary operator that encodes the dual cycle orientations around the vertex. A \emph{regular} vertex $v$ has $I_v=0$, and otherwise it is called \emph{singular}. The field on the $1$-ring of a regular vertex can be \emph{combed} (see Figure~\ref{fig:matching-combing}): it can be \emph{locally} re-indexed  in every face of the $1$-ring such that $\forall e \in N(v),\ I_e=0$. With re-indexing, an $N$-field is locally reduced to $N$ independent fields. A \emph{fractional} singular vertex is defined by having $I_v \notin \mathbb{N}$, where such combing is not possible. Fields with fractional singularities cannot be \emph{globally} combed. This is generally the case, as $\sum_{\forall v \in V}{I_v}=\chi\left(\mathcal{M}\right)$, with $\chi\left(\mathcal{M}\right)$ the Euler characteristic of the mesh. \emph{Integral singularities} do not induce matching mismatches, and therefore appear in single-vector fields as well, as sources, sinks, and vortices. They are basically sources of divergence and curl, and are irrelevant to our extension to $N$-directional fields.
\subsection{Extending FEM calculus} 

\begin{figure}
	\centering
	\includegraphics[width=0.5\textwidth]{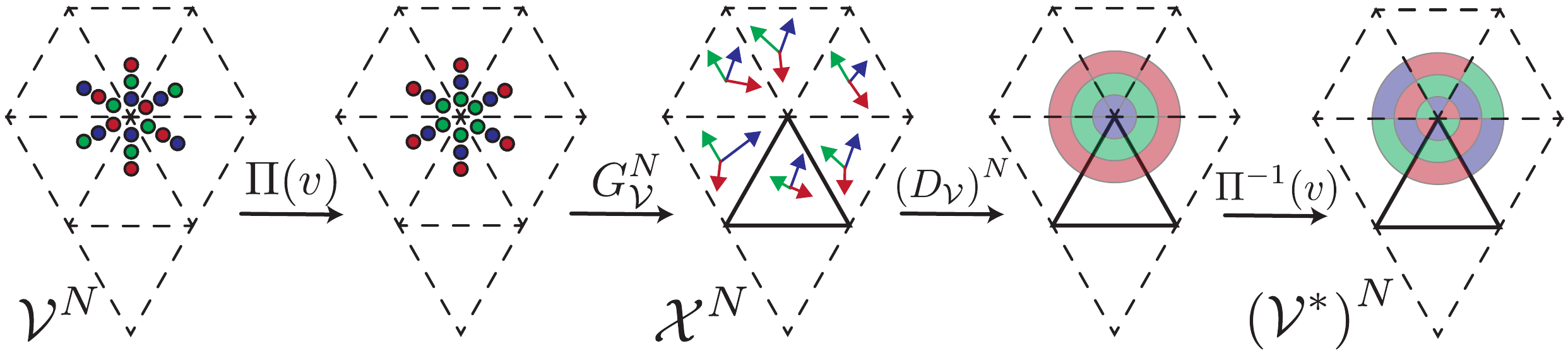}
	\caption{Example of directional calculus with the $N$-Laplacian on a regular vertex. A corner-based vertex function is combed to a single vertex function, after which a gradient is applied to obtain a combed directional field. Applying the divergence then results in a combed integrated value on the vertex, which is combed back to the original labeling.}
	\label{fig:directional-calculus}
\end{figure}

To be able to extend our subdivision scheme for $N$-directional fields, we need a concept of $N$-halfedge forms, $N$-scalar functions, and the entire suite of differential operators. For this, we next adapt existing notions from discrete calculus of branching coverings~\cite{Kalberer:2007,Bommes:2009,Diamanti:2015}. See Figure~\ref{fig:directional-calculus} for an exemplification of the directional calculus presented here.

\paragraph{Seamless function spaces}
Consider a vertex $v \in V$ with adjacent faces (in CCW order) $t_1,\ldots, t_d$, and associated corners $v_1, \ldots, v_d$. Further consider edges $e_i$ between corners $v_i$ and $v_{i+1}$. The function space $\mathcal{V}^N$ is parameterized by a vector $\mathbf{f}_{v_i}$ of $N$ functions per corner $i$: $\mathbf{f}_{v_i} = \left(f_{v_1},\ldots, f_{v_N}\right)^T$. This amounts to $N \cdot d$ values for a single vertex (they are in fact spanned by a lower-dimensional parameter space, as we see in the following). The functions are matched across edges similarly to $N$-directional fields: consider two adjacent corners $v_i$ and $v_{i+1}$ across edge $e_i$ with matching index $I_{e_i}$. We construct the permutation matrix $\pi(e_i)$ that represents the map that the matching induces, to obtain:
\begin{equation}
\mathbf{f}_{v_{i+1}} = \pi(e_i) \cdot \mathbf{f}_{v_i}
\label{eq:function-matching}
\end{equation}

We always assume that within a single face, the corners have a trivial matching (so they are separate $N$ functions); the only non-trivial matching is between corners across edges.

\paragraph{Combing}

For regular vertices, and by successively applying Equation~\ref{eq:function-matching}, we get that $\mathbf{f}_{v_1} = \pi(e_d) \cdot \mathbf{f}_{v_d}$. As such, we can \emph{comb} the functions over regular vertices, in the same way we do for directional fields: for a single $1$-ring, we start from corner $v_1$ in face $t_1$, and transform every $\mathbf{f}_{v_i}$ into $\mathbf{f}_{v_1}$ by inverting Equation~\ref{eq:function-matching} recursively. We denote this linear transformation as $\Pi(v)$. Note that it means that there are only $N$ independent functions in every regular vertex, parameterized by $\mathbf{f}_{v_1}$, which is expected.

\paragraph{Conforming operators}

All the conforming differential operators can be directly extended from the single-vector calculus around regular vertices, by conjugation with the combing (see Figure~\ref{fig:directional-calculus}). For instance, we have that the divergence $D^N(v):\mathcal{X}^N \rightarrow \mathcal{V^{\ast}}^N$ is:

\begin{equation}
D^N = \Pi^{-1}(v) \begin{pmatrix} d_0^T M_\mathcal{X} & & \\  & \ldots & \\ & &  d_0^T M_\mathcal{X} \end{pmatrix}\Pi(v).
\end{equation}

In words, we comb a function and a field around a regular vertex, use the operators on every function in the vector $f_{v_1}$ independently, and then comb back. The result is a vector of $N$ scalars representing the independent divergences of the combed functions. Then, $\Pi^{-1}(v)$ combs the $N$ scalars to corner-based values corresponding to original corner indexing. It is important to note that the identity of the ``first'' corner $v_1$ does not result in any loss of generality, due to the conjugation with $\Pi(v)$; the result per corner would be exactly the same regardless of which corner is first.

The gradient operator $G_V$ extends to $G^N_\mathcal{V}:\mathcal{V}^N \rightarrow \mathcal{X}^N$ by simply operating on the elements in the function values of the corners of the face independently, to produce $N$ vectors. Therefore it doesn't require combing; the corners of every single face are always trivially matched to each other.

%It is important to note that the identity of the first corner $v_1$ changes the output of the differential operators up to a single permutation per vertex. However, as long as the identity of the corner is fixed, the differential operators are well-defined, where the specific identity of the first corner only permutes the values in the result. Therefore, there is no loss of generality.

\paragraph{Non-conforming operators}

Non-conforming differential operators, namely the curl $C^N$, are easier to generalize: we only have to locally comb two faces sharing a single edge, and then conjugate the curl operator independently for the $N$ vectors in both faces with the combing operation. The result is a function in $\mathcal{E^{\ast}}^N$. The rotated co-gradient $JG_\mathcal{E}^N$, exactly like $G^N_\mathcal{V}$, is defined per-face and therefore does not require any matching or combing. %\BAC{does it matter that the edge field is somewhat ''shared'' between the faces?}

\paragraph{Structure-preserving calculus}

It is easy to verify that directional-field calculus is structure-preserving with relation to the sequence around regular vertices. We have that $C^N\cdot G^N_\mathcal{V} = 0$, and that \mbox{$D^N\cdot J\cdot G^N_\mathcal{E}=0$} as well. The formal proof is straightforward, given the conjugation of combing and differential operators, and we omit it for brevity. Essentially, the existence of such sequences means that we can also define a directional Hodge decomposition, but we leave this line of research for future work.

\paragraph{Around singular vertices}
For singular vertices, the product of $\pi(e)$ matrices leads to a non-trivial permutation matrix. That is, ``returning'' to $v_1$ after applying Equation~\ref{eq:function-matching} successively, we get $\mathbf{f}_{v_1} \neq \pi(e_d) \cdot  \mathbf{f}_{v_1} $. As such, conforming differential operators are not well-defined for fractional singularities. To rationalize this, they can be interpreted as isolated boundary points in the field where there is not enough continuity by definition to allow for well-defined conforming operators. The non-conforming operators are well defined everywhere, as they only require two faces in every stencil.

\subsection{Extending $\Gamma^N$} 

Calculus of halfedges is natural in the $N$-directional setting. We define $\mathbb{\gamma} \in \Gamma^N$ as a vector of $N$ scalars per halfedge. The operators $d_{0,\Gamma}^N$ and $d_{1,\Gamma}^N$ are trivially extended with respect to the matching of the corners. Note that we have a null-sum constraint for each element of $\gamma$ independently. The same is done for per-face operators $P^N:\mathcal{X}^N\rightarrow \Gamma^N$  (and its inverse), unpacking operator $U^N$, and the summation operator $A_{\mathcal{E^{\ast}}^N \rightarrow \mathcal{F^{\ast}}^N}$.

The mean-curl representation, and consequently the operator $W^N$, are defined with the combing in the same manner as nonconforming differential operators like $C^N$: one of the halfedges in every edge is chosen arbitrarily as the ``first'', and then we define $A_{\Gamma^N \rightarrow \mathcal{Z}_1^N}$ to conjugate with the matching. As such, both the resulting mean $z_1$ and (half) curl $\epsilon$ are defined with relation to one of the halfedges, and this choice of ``first halfedge'' is well-defined up to permutation.

\subsection{Extending subdivision operators}

Equipped with an extension of the $\Gamma$ representation to $\Gamma^N$, we next extend our subdivision operators to work with directional fields and preserve their structure.

\paragraph{Branched Loop and half-box splines}

For regular vertices, both the Loop $S_\mathcal{V}$ and the half-box spline $S_\mathcal{F}^{\ast}$ subdivision operators extend to the branched spaces $\mathcal{V}^{N}$ and $\left(\mathcal{F}^{\ast}\right)^N$ by conjugation with combing as well. For instance, for Loop subdivision we get:
\begin{equation}
\left(S_\mathcal{V}\right)^N = \Pi^{-1}(v) \begin{pmatrix} S_\mathcal{V} & & \\ & \ldots &  \\ & & S_\mathcal{V} \end{pmatrix}  \Pi(v).
\end{equation}

The result creates new even and odd edges, where the permutation $\pi(e)$ for even edges is the same as the coarse edges they originate from, whereas $\pi(e)$ for odd edges is an identity, since they are created within coarse faces.

\begin{figure}
	\centering
	\includegraphics[width=0.5\textwidth]{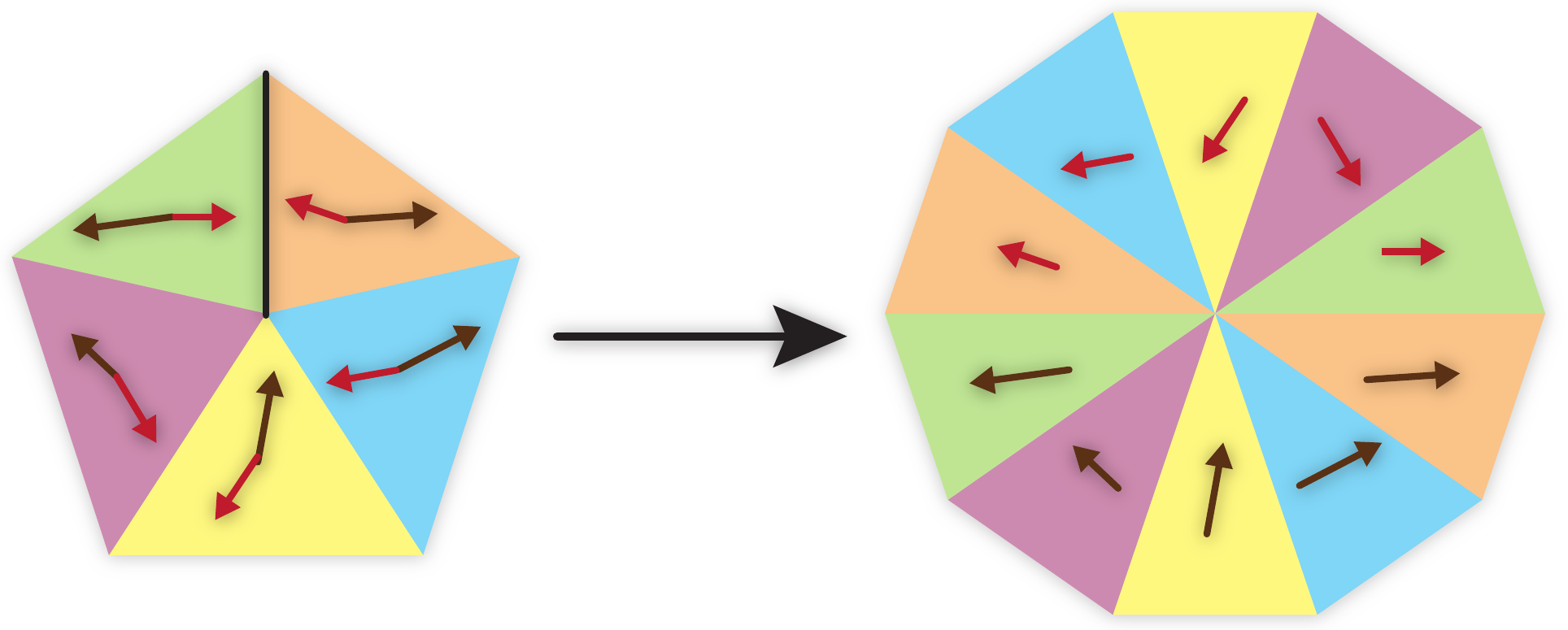}
	\caption{The unfolding operator $\Phi(v)$, illustrated for a singular vertex in the space $\mathcal{X}^N$. We unfold a valence $5$ vertex with singularity index $-\frac{1}{2}$ into a valence $10$ ring with a single vector field. The vector field is then locally subdivided with $S_\Gamma$ and then folded back.}
	\label{fig:unfolding}
\end{figure}

\begin{figure}
	\centering
	\includegraphics[scale=1]{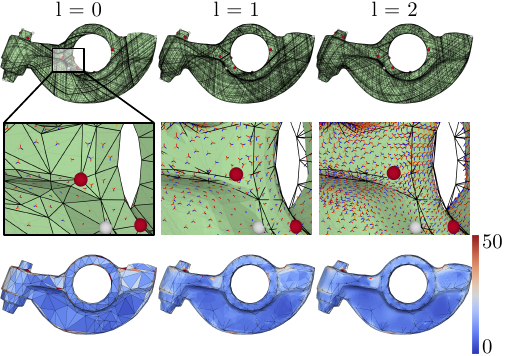}
	\caption{A subdivided $3$-directional field. The spheres mark the singularities (red = $-\frac{1}{3}$, white = $\frac{1}{3}$). Middle: zooming in around a singularity. The color-coding shows the vector norm of the curl per edge, averaged to faces and divided by face area. The curl is evidently refined under subdivision.}
\label{fig:directional-rocker}
\end{figure}

\begin{figure}
	\centering
	\includegraphics[scale=1]{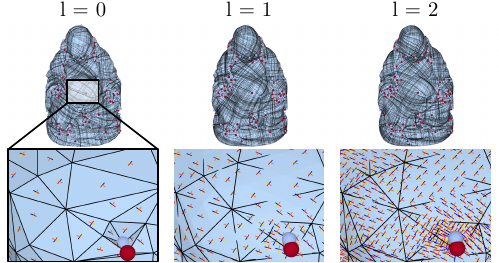}
	\caption{An approximately curl-free $4$-directional field subdivided. Bottom: zoom-in, with singularity colors red = $-\frac{1}{4}$, orange = $-\frac{2}{4}$, and white = $\frac{1}{4}$.  The $L_\infty$ norms of the curl per edge for the three levels are 5.05e-6, 9.50e-7 and 2.18e-7 respectively, which demonstrate that the (lack of) curl of the field is preserved under subdivision.}
		\label{fig:directional-buddha}
\end{figure}

For singular vertices, we require a different definition of the subdivision operators. We do so by \emph{unfolding the branch} (see Figure~\ref{fig:unfolding}): consider again a one ring with $d$ faces, with singularity index $I_V = \frac{i}{N}$. We pick a single vector, follow its matching around the ring until we reach it again, and create a new ring just with this vector. We then do so until all vectors are taken. That creates $\text{GCD}(i, N)$ (greatest common divisor) new rings. We are always guaranteed to return to the original vector since $\left(\pi(v)\right)^N = I$. We denote the unfolding operation as $\Phi(v)$. Then, we can conjugate $S_\mathcal{V}$ for singular vertices with the unfolding:
$$
\left(S_\mathcal{V}\right)^N = \Phi(v)^{-1} S_\mathcal{V}  \Phi(v).
$$
The unfolding $\Phi(v)$ is a generalization of the combing operator $\Pi(v)$ that allows us to extend all our subdivision operators without altering the original scalar subdivision stencils, as the commutation also works through the conjugation. For example, for a regular vertex we just create $N$ new rings each with the separated single vector field. As a result, we maintain all the differential properties of the subdivision, and among them structure-preserving of curl and exactness. We demonstrate this in Figures~\ref{fig:directional-rocker} and~\ref{fig:directional-buddha}.

% !TEX root =  SubdivisionDirectionalFields.tex

\section{Applications}
\label{sec:applications}

In the following, we apply our SHM framework to several applications that use PCVF directional fields in their pipeline. We implemented the subdivision operators in C++ using the Directional library~\cite{Directional}, and the applications using MATLAB. Times are measured on a laptop with an Intel i7-7700HQ (2.8GHz) CPU and 32 GB of RAM. 

\paragraph{Vector field design}

\begin{figure}[h!]
	\centering
	\includegraphics[width=0.5\textwidth]{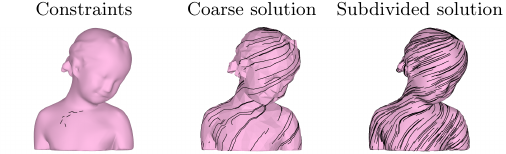}
	\caption{SHM Vector-field design. The local constraints (left) are interpolated to the rest of the coarse mesh (center), and subdivided to the fine mesh (right) at $l=3$.}
	\label{fig:vector-field-design}
\end{figure}

In Figure~\ref{fig:vector-field-design} we show an example of coarse-to-fine vector field design. Vectors are constrained on a small set of faces of a coarse mesh, and interpolated to the rest of the mesh by minimizing the SHM (with level $l=3$) Hodge energy:
$$E_H \left(\gamma\right) = \mathbb{M}_{\Gamma^0} \left(\left| \mathbb{C}_\Gamma \gamma^0 \right|^2 +\left| \mathbb{D}_\Gamma \gamma^0 \right|^2\right)$$

of a field $\gamma^0$ on the coarse mesh. This is done by solving $\mathbb{L}^3_{\Gamma}\gamma^0 = 0$ with fixed values for a subset of constrained faces. We then subdivide $\gamma^0$ to get $\gamma^3$ as our result. We get a fine smooth field efficiently designed with the coarse (restricted) degrees of freedom.

\paragraph{Earth mover's distance}

We apply our subdivision to the optimal transport algorithm presented in~\cite{Solomon:2014}. For brevity, we do not consider meshes with boundary in this experiment. The formulation computes a geodesic vector field between two probability distributions $\mu_0, \mu_1 \in \mathcal{V^{\ast}}$ with $\sum_{v\in V}{\mu_{0|v}} =\sum_{v\in V}{\mu_{1|v}} = 1$, which are defined on the fine mesh of level $l$. These distributions are defined by densities $\rho_{0|1} = M_\mathcal{V}^{-1} \mu_{0|1} \in \mathcal{V}$. The geodesic field is computed to minimize (a simplification of) the 1-Wasserstein distance $\zeta(\mu_0, \mu_1)$ between the probability measures as follows:
\begin{align}
\zeta(\mu_0, \mu_1) &= \inf_{g,h}{\sum_{t \in F^l}{A(t)\left| G_{\mathcal{V}|t}f + JG_{\mathcal{E}|t}g + h \right|_{L_2}}} \\ \nonumber
s.t.\ \ L^l_\mathcal{V}f &= M^l_\mathcal{V}(\rho_0-\rho_1),
\end{align}
where $g \in \mathcal{E}^l$ and $h$ is a harmonic field in $H^l$. $f \in \mathcal{V}^l$ is fully determined from the Laplacian constraint. To limit the solution space on a fine mesh, they use a spectral subspace for $g$ from its Laplacian $L_\mathcal{E}$. We offer an alternative low-rank SHM approximation that uses coarse-mesh function values instead, which is more efficient due to the sparsity of the subdivision matrix. Here, we deviate from the multigrid $V$-cycle folding paradigm of SHM, and solve the problem directly on the fine mesh. Nevertheless, we limit the solution space to subdivided coarse functions. To use the refinable conforming functions, we note that the underlying continuous norm is invariant to rotations. Therefore, we dualize the discretization of the problem: we consider mid-edge distributions $\rho'_0, \rho'_1 \in \mathcal{E}^{\ast}$, transform the problem to refinable $\gamma \in \Gamma$, and solve for:
\begin{align}
\zeta(\mu_0, \mu_1) &= \inf_{f^0,h}{\sum_{t \in F^k}{\sqrt{\left(\gamma^l_{|t}\right)^T\cdot M^l_{\Gamma}\cdot \gamma^l_{|t}}}} \\ \nonumber
s.t. \ \gamma^l &= S_\Gamma \left(d_{0,\Gamma}f^0\right) + JG^l_\mathcal{E}g^l + S_\Gamma h^0 \\ \nonumber
\ \ L^l_\mathcal{E}g^l &= M^l_\mathcal{E}\left(\rho'_0-\rho'_1\right).
\end{align}
In words, we solve for coarse $f^0$ so that its subdivided gradient $\gamma^l$, creates the least-norm vector field with the Laplacian-computed coexact component $g$ (we use a simply-connected mesh with no harmonic component for simplicity). This is solved using the ADMM procedure described by~\cite{Solomon:2014}. Note that the coexact component is computed beforehand, and therefore fixed after solving the Laplacian equation.

\begin{figure}[h!]
	\centering
	\includegraphics[scale=0.8]{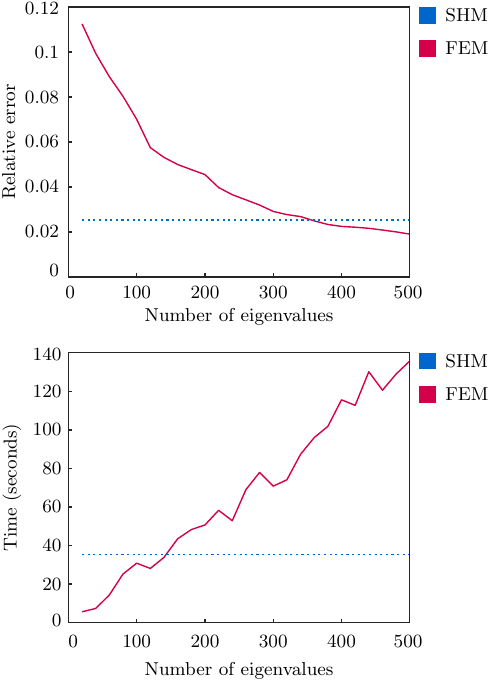}
	\caption{Top: relative error $\nicefrac{\left|\zeta-\zeta_{GT}\right|}{\zeta_{GT}}$ of the computed distance value via SHM (horizontal dotted line, for reference) and the FEM spectral subspace approximation with number of eigenvalues from 20 to 500 in steps of 20. 
	Bottom: total running time for SHM (horizontal dotted line, for reference) and the spectral approximation for the above specified number of eigenvalues.}
		\label{fig:emd-error-and-time}
\end{figure}

\begin{figure}[h!]
	\centering
	\includegraphics[width=0.5\textwidth]{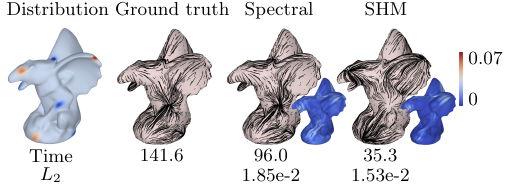}
	\caption{A Comparison of the EMD algorithm between the spectral (center-right) approximation with 360 eigenvalues, and our SHM (right), with similar accuracy with relation to the ground truth (center-left). Left: the initial mass distribution. The big figures depict the geodesic fields $\gamma^l$ resulting from the optimization, with insets depicting the vector norm per face of the difference between the fine level solution and the chosen approximation. The total running times are given: for the spectral-subspace solution, 68.1 seconds were spent on the ADMM optimization and 27.8 seconds on computing the basis and prefactoring. For SHM, this was 30.8 and 4.5 seconds respectively. The $L_2$ error to the ground truth is provided.}
		\label{fig:emd}
\end{figure}

Our experiment is conducted as follows: we compute our SHM solution, and compare the result to a spectral-subspace FEM solution with an increasing number of eigenbases. A similar accuracy (measured to the ground-truth solution in the fine level) is achieved with approximately $360$ eigenvalues, at almost three times the computation time. We show our results in Figures~\ref{fig:emd-error-and-time} and Figure~\ref{fig:emd}.
%For a fair comparison, we use the same amount of fine eigenfunctions (their method) as the number of coarse vertices (in our formulation), so that the problem is solved with the same dimensionality. While the eigendecomposition provides a slightly better accuracy in the EMD distance itself, our timing (and memory complexity) are superior. 


\paragraph{Operator-based advection}

Our framework can be used to modify the operator-based representation of PCVFs introduced in~\cite{Azencot:2013, Azencot:2015}. Their method constructs a discrete version of the classical representation of vector fields as derivations of scalar functions: for a field $u$ and a scalar function $f$, the operator produces $\nabla_uf=\langle u, \nabla f\rangle$. Given a vector field $u \in \mathcal{X}$, their discrete operator is represented by a matrix $B_\mathcal{V}: \left|V\right| \times \left|V\right|$ on a mesh that is composed as follows:
$$
B_\mathcal{V} = \frac{1}{3}\left(M_{\mathcal{V}}\right)^{-1}A_{\mathcal{F} \rightarrow \mathcal{V}} M_\mathcal{X}B_\mathcal{F}G_\mathcal{V},
$$
where $B_\mathcal{F}: \left|F\right| \times 3\left|F\right|$ is a matrix that performs the facewise dot-product of the face-based gradient with $u$, and $A_{\mathcal{F} \rightarrow \mathcal{V}}$ sums values from faces to adjacent vertices, in our usual notation. Essentially, the dot products are made per face, and averaged to the vertices using the respective mass matrices of the mesh.

The operator representation makes it simple to advect a function $f \in \mathcal{V}$ on a surface: given time $t$, and the initial function value $f(0)$, they solve the advection equation in the weak sense, integrated over a spectrally-reduced subspace. Consider $\Psi$ as the matrix with $n$ lowest Laplacian eigenvectors as columns. They work with a vector of coefficients $\alpha$ so that $f(t)= \Psi \alpha(t)$, and solve for:
$$
\Psi^T M_\mathcal{V}  \cdot \frac{\partial \left(\Psi \alpha(t)\right)}{\partial t} = \Psi^T M_\mathcal{V} \cdot B_\mathcal{V}\Psi \alpha(t), 
$$
Where $\Psi^T M_\mathcal{V}\Psi=Id$. We get $\alpha(t) = \exp\left(t \cdot B_\mathcal{V}^{EIG}\right)\cdot \alpha(0)$, where 
$$
B_\mathcal{V}^{EIG} =  \Psi^T M_\mathcal{V}B_V \Psi.
$$
We follow a similar construction, except that we integrate over a subspace of refined subdivision basis functions, and our degrees of freedom are the coarse function $f^0$ so that $f(t) = S_{\mathcal{V}} f^0(t)$. Posing the system in the weak sense, we get:
$$
S_\mathcal{V}^T M_\mathcal{V} \cdot \frac{\partial S_\mathcal{V} f^0(t)}{\partial t} = S_\mathcal{V}^T M_\mathcal{V} \cdot B_\mathcal{V}S_\mathcal{V} f^0(t)
$$
where the solution is $f^0(t) = \exp\left(t \cdot B_\mathcal{V}^{SHM}\right)\cdot f^0(0)$, and 
$$
B_\mathcal{V}^{SHM} = \mathbb{M}_\mathcal{V}^{-1} S_\mathcal{V}^T M_\mathcal{V}B_\mathcal{V}S_\mathcal{V}.
$$
The weak form is natural to the eigenfunction reduction, as the eigenfunctions are orthogonal w.r.t. $M_\mathcal{V}$, and $\psi^{\dagger}=\psi^TM_\mathcal{V}$. Nevertheless, we empirically witnessed that omitting $M_\mathcal{V}$ actually slightly improves the accuracy of SHM advection; we conjecture that this is since the subdivision bases are ``more orthogonal'' w.r.t. a uniform matrix, but reserve the concrete analysis for future work.

%They  as well: define $\Psi$ as the matrix with the Laplacian eigenvectors as columns, then the eigenfunction approximation of $B_\mathcal{V}$ is as follows:
%$$
%B_\mathcal{V}^{EIG} = \Psi^{-1}B_\mathcal{V}\Psi = \Psi^T M_\mathcal{V}B_V \Psi.
%$$

%To make the operator-based approach \changed{SHM}-compatible, we use the following formulation instead:
%$$B_\mathcal{V}^{\changed{SHM}} = \left(\mathbb{S}_\mathcal{V}\right)^{\changed{+}} B_\mathcal{V} \mathbb{S}_\mathcal{V},$$

%where the $B_\mathcal{F}$ operator that is part of $B_\mathcal{V}$ is made of a subdivided coarse vector field. \changed{This form in fact solves the advection equation for subdivided functions in a weak form (integrated against $\left(\mathbb{S}_\mathcal{V}\right)^T$):
%\begin{align}
%\left(\mathbb{S}_\mathcal{V}\right)^T \cdot \frac{\partial \mathbb{S}_\mathcal{V}f(t)}{\partial t} &= \left(\mathbb{S}_\mathcal{V}\right)^T \cdot B_\mathcal{V}\mathbb{S}_\mathcal{V}f \Rightarrow \\
%f(t) &= exp\left(t \cdot  \left(\mathbb{S}_\mathcal{V}\right)^{+} B_\mathcal{V} %\mathbb{S}_\mathcal{V}\right)f(0),
%\end{align}

%\changed{as by definition $\left(\mathbb{S}_\mathcal{V}\right)^{+} =\left(\left(\mathbb{S}_\mathcal{V}\right)^T \cdot \mathbb{S}_\mathcal{V}\right)^{-1} \left(\mathbb{S}_\mathcal{V}\right)^T$}. 

\begin{figure}
	\centering
	\includegraphics[width=0.5\textwidth]{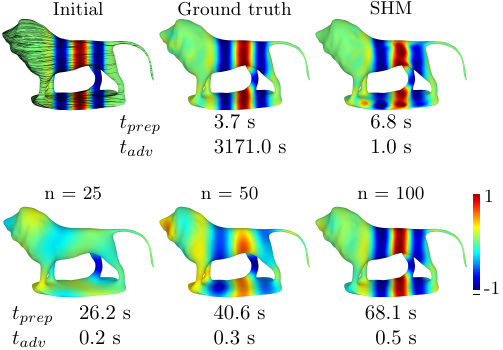}
	\caption{Operator-based advection. The initial (color-coded) function $f$ is projected to the reduced basis and advected with the given vector field $v$ (streamlines), with time $t=0.5$. The bottom row shows the approximation using the first $n$ Laplacian eigenfunctions. We provide the preparation and advection times $t_{\text{prep}}$ and $t_{\text{adv}}$. $v$ is given by $v_t(x,y,z) = -x$ per triangle $t$ where $x,y,z,$ are the coordinates of the barycenter of $t$. $f_0$ (before projection) is given by $f_{0,v}(x,y,z) = \sin(\frac{4}{5}\pi x)$ for the $x$ coordinate of vertex $v$. We localize the function below  $x_{min}=\min(V_x) + \frac{1}{3}(\max(V_x)-\min(V_x))$ and above $x_{max}=\min(V_x) + \frac{2}{3}(\max(V_x)-\min(V_x))$ by multiplying $f_{0,v}$ with $\exp(-5|v_x-x_{min}|)$ below $x_{min}$ and $\exp(-5|v_x-x_{max}|)$ above $x_{max}$. The model has $|V^0| = 435$ and $|V^5|=449532$.}
	\label{fig:operator-based-advection}
\end{figure}

\begin{figure}
	\centering
	\includegraphics[width=0.5\textwidth]{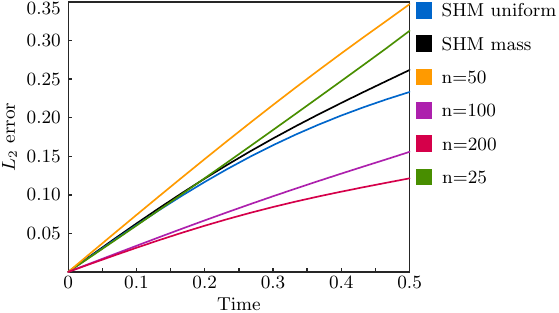}
	\caption{Normalized $L_2$ error for the SHM (with and without mass matrix $M_\mathcal{V}$) and spectral approximations, depicted in Figure \ref{fig:operator-based-advection}, compared to the ground truth for times $t=[0,0.5]$, with step size of $\Delta t = \frac{1}{32}$. We define the ground truth $f(t)$ as the initial function, expanded in the respective basis and advected with the fine-level operator.} %We then calculate the normalized $L_2$ error at the fine level as $\sqrt{(f(t)-\bar{f}(t))^T M_\mathcal{V} (f(t)-f^*(t)) / (f^*(t)^T M_\mathcal{V} \bar{f}(t))}$}}
	\label{fig:advection-error-graph}
\end{figure}

We compute a ground-truth solution at the fine level $l=5$, project it to the reduced basis, and compare the SHM advection (both uniform and weights with $M_\mathcal{V}$) against the spectral advection for a different number of eigenbases. We show the result in Figure~\ref{fig:operator-based-advection}, and the error in Figure~\ref{fig:advection-error-graph}. The spectral-subspace approximation has a comparable error profile between $50$ and $100$ eigens, but the computation is about 8--10 times as slow, where the eigenbasis extraction is the expensive part. Note that For both SHM and the spectral approximation, the error diverges with time, due to the high frequencies inevitably created by the advection equation.

%\changed{From the timings in Figure~\ref{fig:operator-based-advection}, we can conclude that when performing more iterations, the lower dimensional approximations start to outperform the fine level computation. Among the approximations, the setup time for SHM is far less than the approximation with 300 eigenvalues that gives comparable error with respect to the ground truth.}\BAC{discuss timings more?}

% analyzes the error for different eigenvector resolutions. There we see that for the given time range, the \changed{SHM} solution performs better compared to the <200 eigen functions approximations, despite the use of the dense pseudo-inverse $S_\mathcal{V}^{-1}$. 


\paragraph{Seamless parameterization}

\begin{figure}[h!]
\centering
\includegraphics[width=0.5\textwidth]{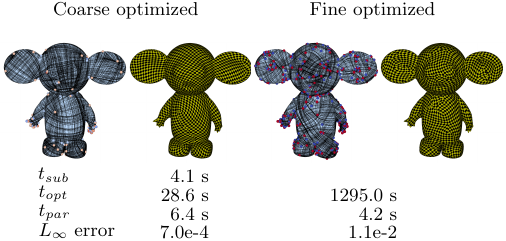}
\caption{Seamless parameterizations with coarse-to-fine curl-free fields. Left: coarse curl optimization and subdivision, and fine-level parameterization. Right: fine curl optimization and parameterization. The run time is significantly reduced by optimizing in the coarse level, the integration error is lower, and the result is more appealing. The fine level is at $l=2$. We use 250 iterations for the curl optimization. $t_{sub}$ is the subdivision time, $t_{opt}$ is the curl optimization time and $t_{par}$ is the parameterization time. The  optimization time evidently dominates the total running time.
}
\label{fig:seamless-parameterization}
\end{figure}

We next employ our structure-preserving subdivision for $N$-directional fields to compute coarse-to-fine curl-reduced fields. This allows us to compute fine-level rotationally-seamless parameterizations (where the direction identifies across cuts, but without perfect integer translations) with a very low integration error. We compute an $N$-RoSy (with~\cite{Knoppel:2013}) on the coarse mesh, optimize it to be (approximately) curl-free with~\cite{Diamanti:2015}, and compute a coarse paramaterization that consequently has a very small integration error. The subdivision preserves the small amount of curl, and the fine-level parameterization also has a small error as a result. We compare this process to performing the curl-free optimization on the fine mesh directly. Our coarse-to-fine optimization is faster in almost two orders of magnitude. We demonstrate this in Figure~\ref{fig:seamless-parameterization} and in the teaser (Figure~\ref{fig:teaser}).

% !TEX root =  SubdivisionDirectionalFields.tex

\section{Discussion}
\label{sec:discussion}

\paragraph{Convergence and smoothness} As we discuss in the auxiliary material, our subdivision stencils for $S_\mathcal{E^{\ast}}$ and $S_\Gamma$ have a few degrees of freedom (after counting the commutation constraints) that we use to optimize the spectrum of the subdivision stencil, such that it converges in the limit since the subdominant eigenvalues are less than $1$. We conjecture that since the fields are derivatives of smoothly-subdivided functions, they are then one level of smoothness lower in the limit. However, we leave a formal theoretical analysis of convergence and smoothness to future work. We believe that a better design practice might be to allow $S_\mathcal{V}$ and $S_\mathcal{F^{\ast}}$ to vary entirely, where the smoothness of all subdivision operators is optimized concurrently (similarly to~\cite{Huang:2010}), rather than modify the existing schemes.

\paragraph{Dual formulation} Our $\Gamma$ space uses $\langle v, e \rangle$ for the projection operator $P$. Nevertheless, the entire formulation can be made with the perpendicular $\langle v, e^{\perp} \rangle$, using the non-conforming divergence and conforming curl. This can be beneficial to fluid simulation.

\paragraph{Preconditioning and its disadvantages} The mass matrices of SHM are generally more strongly positive-definite than those of the FEM in the coarse mesh. The reason is that the uniform and stationary subdivision operators average the mesh, and create better triangulations. Nevertheless, the fact that the subdivision does not commute with the fine mass matrix also creates the high-frequency divergence pollution in the subdivided fields. It is then worthwhile to try and explore alternatives that consider the mass matrices within the templates, to obtain precise fine Hodge decompositions.

\paragraph{Full multi-resolution processing} Our paper explores low-dimensional coarse-to-fine approximations. Moreover, the basis functions are not orthonormal as an eigenbasis, albeit considerably cheaper to obtain. Nevertheless, SHM can be augmented by incorporating biorthogonal \emph{subdivision wavelets}~\cite{Lounsbery:1997,Bertram:2004}, to obtain exact multi-resolution representation of functions over the fine mesh, with the advantages of increasing locality---this could benefit applications such as solving diffusion problems.

\paragraph{Non-triangular meshes} The space $\mathcal{X}$ is not well defined for non-planar polygonal meshes. Nevertheless, in the spirit of mimetic elements~\cite{bossavit:1998}, the space $\Gamma$, with its null-sum constraint, is well-defined for any polygonal mesh, implicitly defining $\mathcal{X}$. As such, our framework can consider other subdivision operators (such as Catmull-Clark). We will explore this in future work.

\paragraph{General restriction operators} Finally, our setting is currently limited to subdivision surfaces. It could be beneficial to also allow for a multi-resolution setting on general fine meshes using simplification operators (such as quadratic-error-based simplification~\cite{Garland:1997}) as the restriction operators. This should prove challenging as the vertex- and face-based restrictions have to be defined first, but will allow a very general framework for directional-field processing on arbitrary triangle meshes.

\section{Acknowledgements}
\label{sec:acknowledgements}

The authors would like to thank Bettina Speckmann for her support, and furthermore Fernando de Goes, Nilima Nigam, Mirela Ben-Chen, Justin Solomon and Etienne Vouga for helpful discussions.

\bibliographystyle{ACM-Reference-Format}
\bibliography{SubdivisionDirectionalFields}
% !TEX root =  SubdivisionDirectionalFields.tex

\appendix
\section{$P^{-1}$ as inverse of $P$}
\label{app:pinv}
We next show that tangential vector fields are preserved under the operation $P^{-1}\cdot P$. 

Let $e_1,e_2,e_3$ be the CCW oriented edges of a face in CCW order (i.e., $s_i=1 \forall i \in \lbrace1,2,3\rbrace$), and let $\alpha_1, \alpha_2,\alpha_3$ be the angle opposite the corresponding edge. Here, we will interpret $e_i$ as column vectors. Consider a tangential vector field, that is locally defined on a triangle $t$ as $v_t=a e_1 + b e_2$, without loss of generality. Then, 
\begin{equation}
P_{|t} v_t = \begin{pmatrix}
a |e_1|^2 + b e_1 \cdot e_2\\
a e_1 \cdot e_2 + b |e_2|^2
\end{pmatrix}
\end{equation}
Applying $P^{-1}$, we get
\begin{align}
P^{-1}_{|t} P_{|t} v_t =& \frac{1}{2A} \begin{pmatrix} -e_2^{\perp,T} &e_1^{\perp,T}\end{pmatrix}\begin{pmatrix}
a |e_1|^2 + b e_1 \cdot e_2\\
a e_1 \cdot e_2 + b |e_2|^2
\end{pmatrix}\\
=&\frac{1}{2A}(-e_2^{\perp,T} (a|e_1|^2+b e_1\cdot e_2) +\\
& e_1^{\perp,T} (b|e_2|^2+a e_1\cdot e_2))\\
=& \frac{1}{2A}(X + Y)
\end{align}
where we renamed the summed terms in the brackets for convenience.
Note that $2A =|e_1|e_2|\sin\alpha_3$. We can express $e_2$ in terms of $e_1$ via orthogonal decomposition
\begin{equation}
e_2 = |e_2| \left(\cos(\pi-\alpha_3)\frac{e_1}{|e_1|} + \sin(\pi-\alpha_3)\frac{e_1^\perp}{|e_1|}\right)
\end{equation}
which allows us to write
\begin{align}
e_1^\perp =&\frac{|e_1|}{|e_2|}\frac{e_2}{\sin\alpha_3} + \cot\alpha_3e_1\\
=&\frac{|e_1|^2}{2A}e_2 + \cot\alpha_3 e_1\\
=&(\cot\alpha_2+\cot\alpha_3)e_2 + \cot\alpha_3e_1
\end{align}
where we use $\frac{|e_1|^2}{2 A} = \cot\alpha_2 + \cot\alpha_3$. Similarly, 
\begin{align}
- e_1 =& |e_1| \left(\cos\alpha_3\frac{e_2}{|e_2|} + \sin\alpha_3\frac{e_2^\perp}{|e_2|}\right) \iff\\
e_2^\perp =& -\frac{|e_2|}{|e_1|}\frac{e_1}{\sin\alpha_3} - \cot\alpha_3 e_2\\
=& -(\cot\alpha_1+\cot\alpha_3)e_1 - \cot\alpha_3e_2
\end{align}

Working out $\frac{X}{2A}$, we get
\begin{align}
\frac{X}{2A} =& \left((\cot\alpha_1 +\cot\alpha_3)e_1 + \cot\alpha_3 e_2\right) (a\frac{|e_1|^2}{2A} + b \frac{e_1\cdot e_2}{2A})\\
 =& \left((\cot\alpha_1+\cot\alpha_3)e_1 + \cot\alpha_3e_2\right) (a(\cot\alpha_2+\cot\alpha_3) -b \cot\alpha_3)\\
=& \cot^2\alpha_3(a e_1 + a e_2 - b e_1 - b e_2) + a e_1 \\
& +a \cot\alpha_2\cot\alpha_3 e_2 - b \cot\alpha_1\cot\alpha_3 e_1
\end{align}
where we use $\frac{e_1 \cdot e_2}{2A} = - s_1 s_2 \cot\alpha_3$ and $\cot\alpha_1\cot\alpha_2+\cot\alpha_1\cot\alpha_3 +\cot\alpha_2\cot\alpha_3=1$ for the interior angles of a triangle. 
For $\frac{Y}{2A}$, we get
\begin{align}
\frac{Y}{2A} =& ((\cot\alpha_2+\cot\alpha_3)e_2 + \cot\alpha_3e_1)(-a\cot\alpha_3 + b(\cot\alpha_1 + \cot\alpha_3))\\
=& \cot^2\alpha_3(-ae_2 -ae_1 + b e_2 + be_1) + be_2 \\
& -a\cot\alpha_2\cot\alpha_3e_2 + b \cot\alpha_1\cot\alpha_3 e_1
\end{align}

combining gives:
\begin{align}
\frac{1}{2A}(X+Y) =& a e_1 + b e_2
\end{align}
as desired.

To generalize this result for arbitrarily oriented edges, we need to multiply $-e_2^\perp$ and $e_1^\perp$ by their appropriate sign, and sign the $\gamma$ values to be correctly oriented. This amounts to 
\begin{equation}
P^{-1}_{|t} = \frac{1}{2A}\begin{pmatrix}-s_2 e_2^{\perp,T} & s_1 e_1^{\perp,T}\end{pmatrix} \begin{pmatrix}s_1&0\\0&s_2\end{pmatrix} = \frac{s_1s_2}{2A}\begin{pmatrix}-e_2^\perp\\ e_1^\perp\end{pmatrix}
\end{equation}
as stated before.

\section{Inner Product on $\Gamma$}
\label{app:m-gamma}

In the following, we develop the inner product mass matrix $M_\Gamma$, to prove the formulation of Equation~\ref{eq:m-gamma}.

Consider a face $t=123$ with three edges $e_1,e_2,e_3$ that are, without loss of generality, positively oriented towards the face. Further consider two halfedge forms $\gamma_x,\gamma_y \in \Gamma$ restricted to the face on these edges: $\gamma_{x|(1,2,3)}$ and $\gamma_{y|(1,2,3)}$, representing respective face-based vectors $v_{x|t}$ and $v_{y|t}$. We ``pack'' their representation to edges $1$ and $2$ alone, and by so trivially encoding the null-sum constraint $\gamma_{x|1}+\gamma_{x|2}+\gamma_{x|3}= 0$ (and resp. for $\gamma_y$). Following Equation~\ref{eq:m-gamma-pmp}, we have that the inner product $M_\Gamma$, restricted to the face, is given by:
$$M_\Gamma = P^{-T} M_\mathcal{X} P^{-1}.$$
This reproduces the inner product between $v_y$ and $v_x$ in the face $t$. We then get that:
\begin{align*}
M_\Gamma = P^{-T} M_\mathcal{X} P^{-1} = &\frac{1}{4A_t^2} \begin{pmatrix} -e_2^{\perp} \\ e_1^{\perp} \end{pmatrix} \begin{pmatrix} A_t & & \\ & A_t & \\ & & A_t \end{pmatrix} \begin{pmatrix} -e_2^{\perp} \\ e_1^{\perp} \end{pmatrix}^T = \\
&\frac{1}{4A_t} \begin{pmatrix} e_2^{\perp} \cdot e_2^{\perp} & -e_2^{\perp} \cdot e_1^{\perp} \\  -e_2^{\perp} \cdot e_1^{\perp} & e_1^{\perp} \cdot e_1^{\perp}  \end{pmatrix} 
\end{align*}
Consider the angles $\alpha_{1|2|3}$ opposite to edges $e_{1|2|3}$. Then, we use the identities:
\begin{align*}
\frac{e_1^{\perp} \cdot e_1^{\perp} }{2A_t} &= cot\left(\alpha_2\right) + cot\left(\alpha_3\right)\\
\frac{-e_1^{\perp} \cdot e_2^{\perp}}{2A_t} &= cot\left(\alpha_3\right),
\end{align*}
for any cyclic shift of $\left(1,2,3\right)$. Then we get:
\begin{align*}
M_\Gamma = &\frac{1}{2}\begin{pmatrix} cot\left(\alpha_1\right) + cot\left(\alpha_3\right) && -cot\left(\alpha_3\right) \\ -cot\left(\alpha_3\right) && cot\left(\alpha_2\right) + cot\left(\alpha_3\right) \end{pmatrix} = \\
&\frac{1}{2}U^T \begin{pmatrix} cot(\alpha_1) & & \\ & cot(\alpha_2) & \\ & & cot(\alpha_3) \end{pmatrix} U,
\end{align*}

where we use the unpacking operator $U = \begin{pmatrix} 1 & 0 \\ 0 & 1\\ -1 & -1\end{pmatrix}$.

\pagebreak

\section{Stencils}

\newcommand{\hbSub}{S_{\mathcal{F}^\ast}}
\newcommand{\cSub}{S_{\mathcal{E}^\ast}}
\newcommand{\vSub}{S_\mathcal{V}}
\newcommand{\formSub}{S_1}

In the following we details our subdivision stencils and the way we derived them. We modified the integrated face-based subdivision $\hbSub$ operator, derived from the DEC $S_2$ in SEC~\cite{deGoes:2016sec}, to accommodate for our boundary conditions, and consequently had to modify $S_1$ around boundary vertices. In addition, we introduced a subdivision for unsigned integrated edge functions $\cSub$. We denote the number of incident faces as $d$, so that the regular interior stencils have $d=6$ and the regular boundary stencils have $d=3$. In addition, we denote boundary vertices by a black dot and an interior vertex by an open dot.

\subsection{Loop subdivision} 

\begin{figure}[h!]
\includegraphics[width = 0.5\textwidth]{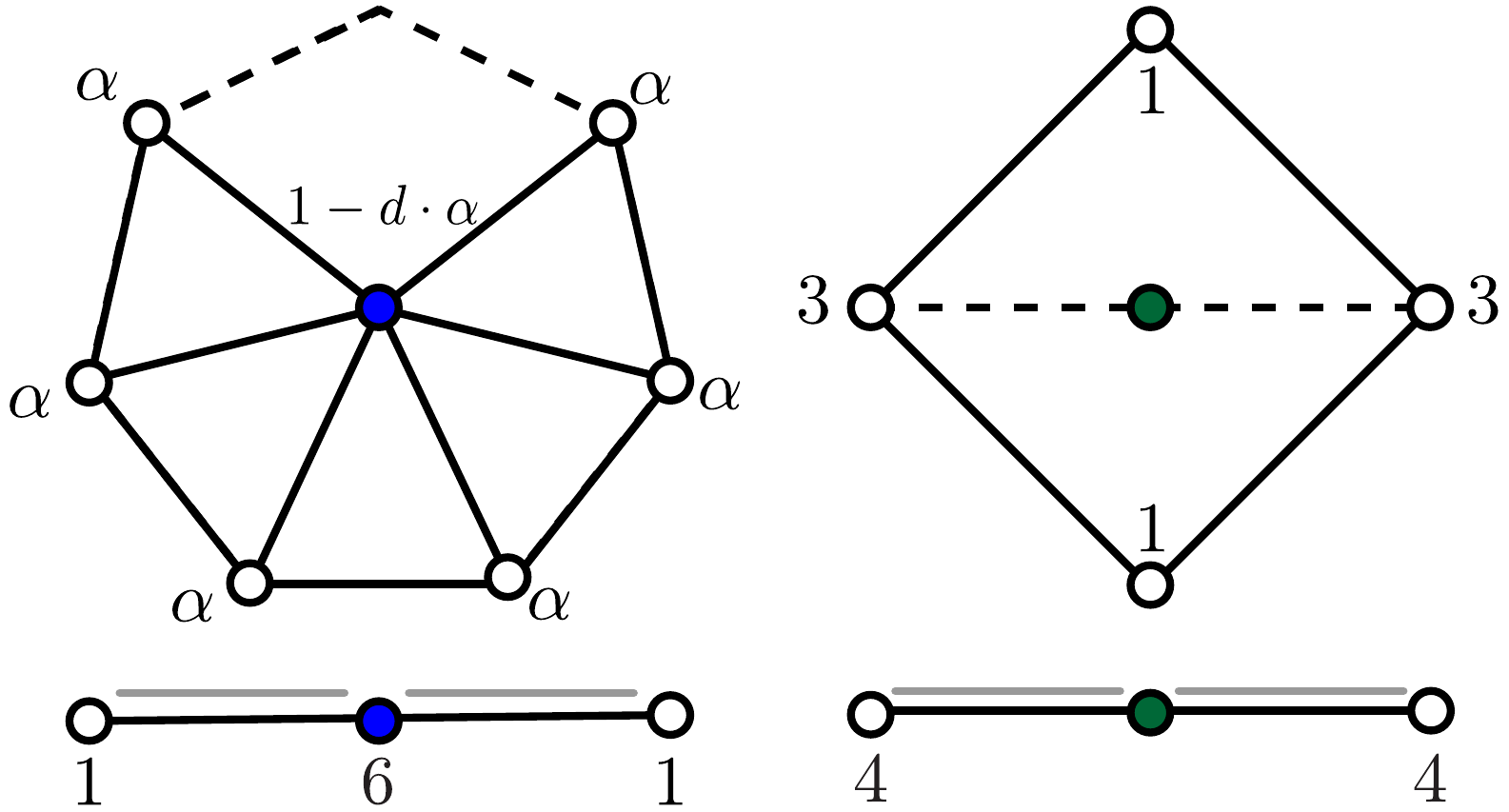}
\caption{Loop subdivision stencils used for $S_\mathcal{V}$ in our setting. The blue dot denotes even vertices (part of the original mesh), whereas the green dots denote the odd vertices (newly inserted in the subdivision step). Double edges denote boundary edges. Multiply all factors by $1/8$, except for the $\alpha$ and $1-d\cdot\alpha$ factors.}
\label{fig:Sv}
\end{figure}

For $S_\mathcal{V}$, we chose Loop subdivision (Fig.~\ref{fig:Sv}) with
\begin{equation}
\alpha = \begin{cases}
	\frac{3}{8d},& d \neq 3\\
	\frac{3}{16}, & d = 3,
\end{cases}
\end{equation}
following Biermann et al.~\shortcite{Biermann:2000}. The templates can be found in Figure~\ref{fig:Sv}. Similar to de Goes et al.~\cite{deGoes:2016sec}, we chose to keep the odd stencil next to the boundary the same as the interior stencil.

\subsection{Halfbox spline subdivision}
\begin{figure}[h!]
\includegraphics[width = 0.5\textwidth]{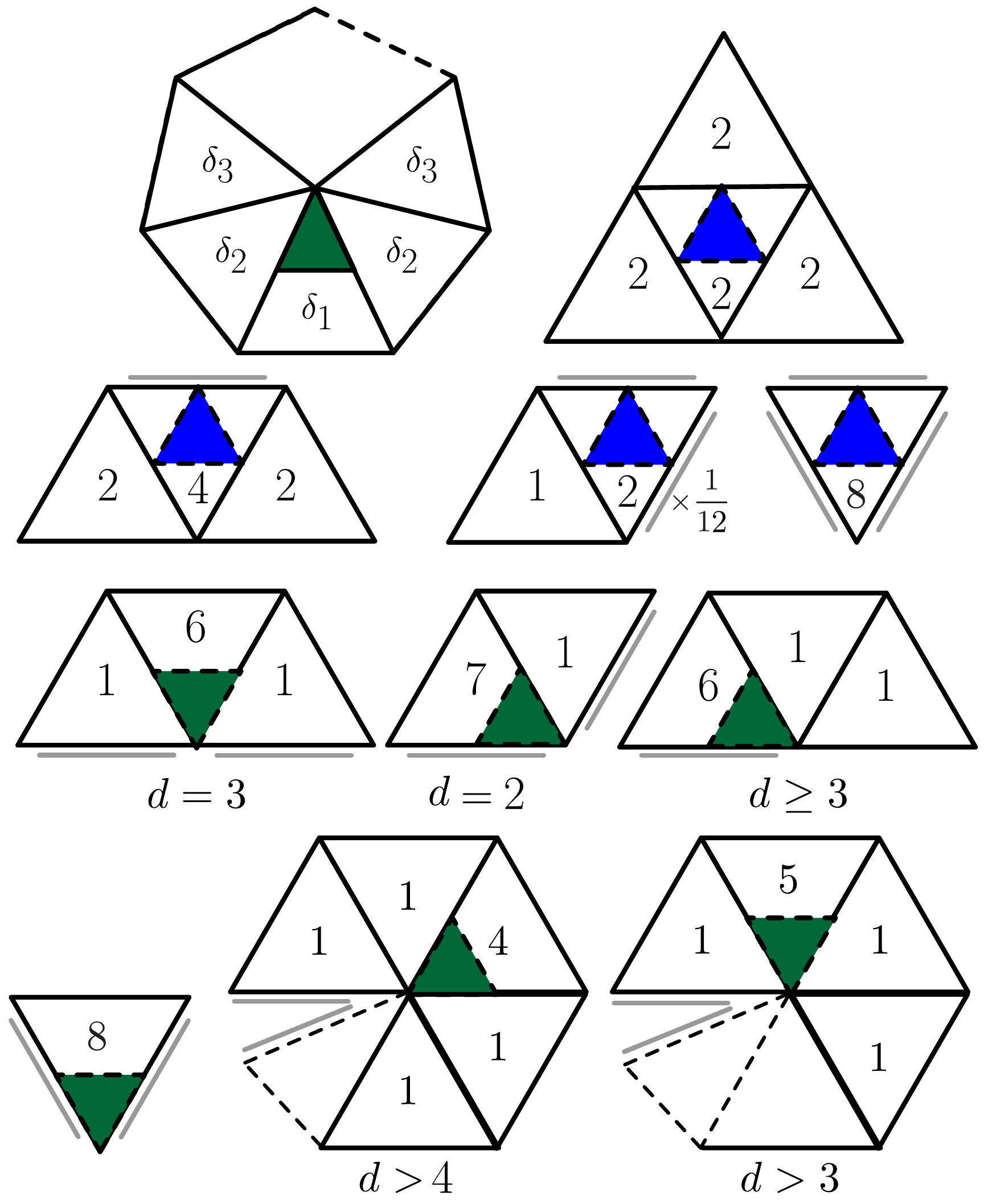}\\
\caption{Modified half-box splines subdivision operator $S_{\mathcal{F}^{\ast}}$. Multiply all factors by $1/32$ except the $\delta_i$ factors or where explicitly stated otherwise. Double edges (with gray) denote boundary edges.}
\label{fig:Sf}
\end{figure}

For the halfbox spline subdivision operator $S_\mathcal{F^{\ast}}$, we use the same stencils as Wang et al.~\shortcite{Wang:2006} for the interior faces, as given by Figure~\ref{fig:Sf}. Due to the extra constraints on $S_{\mathcal{E}^*}$ at the boundary, we modified the boundary stencils for $S_\mathcal{F}$.
The parameters of the interior stencils are given by 
\begin{align}
\delta_1 =& \frac{3}{4} - \beta\\
\delta_2 =& \begin{cases}
\frac{1}{8}& d > 3,\\
\frac{1}{8}+\frac{\beta}{2} & d = 3
\end{cases}\\
\delta_3 =& \begin{cases}
\frac{\beta}{2}& d >4,\\
\beta & d = 4\\
\end{cases}
\end{align}

where $\beta$ is the halfbox spline parameter, given by
\begin{equation}
\beta = \begin{cases}
\frac{1}{12} & d = 3\\
\frac{1}{8} & d= 4\\
\frac{1}{4}-\frac{1}{16}\sin^2\left(\frac{2\pi}{5}\right) & d= 5\\
\frac{1}{4} & \text{otherwise}\\
\end{cases}
\end{equation}

\subsection{1-form subdivision operator}
For completeness, we list the coefficients for the interior stencils of the $S_1$ subdivision operator, as used in deGoes et al.~\cite{deGoes:2016sec}.
\begin{align}
\eta_0 &= \frac{3}{8} - \alpha - \frac{\beta}{4}\\
\eta_1 = \eta_{d-1} &= \begin{cases}
\frac{1}{8} - \alpha + \frac{\beta}{8}& d=3,\\
\frac{1}{8} - \alpha & \text{otherwise}
\end{cases}\\
\eta_2 = \eta_{d-2} &= \begin{cases}
\frac{\beta}{4}-\alpha& d = 4,\\
\frac{\beta}{8}-\alpha& \text{otherwise}
\end{cases}\\
\eta_i &= -\alpha \quad \text{for $2<i<d-2$}\\
\theta_0 = -\theta_{d-1} &= -\frac{\beta}{8}\\
\theta_1 = -\theta_{d-2} &= \begin{cases}
0& d=3,\\
-\frac{\beta}{8}& \text{otherwise}
\end{cases}
\end{align}
with $\alpha, \beta$ the coeffficents for Loop resp. halfbox spline subdivision, as defined before.

\subsection{Stencil Constraints}

The subdivision operators were created with mirror symmetric templates about the target mesh element. The following commutation relations were imposed

\begin{align*}
d_0 S_\mathcal{V} &= S_1 d_1 \\
S_{\mathcal{F}^{\ast}} d_1 &= d_1 S_1 \\
S_{\mathcal{F}^{\ast}} A_{\mathcal{E}^{\ast} \rightarrow \mathcal{F}^{\ast}}  &= A_{\mathcal{E}^{\ast} \rightarrow \mathcal{F}^{\ast}}  {S_\mathcal{E}^{\ast}} \\
C_\Gamma S_{\Gamma}  &= S_{\mathcal{E}^{\ast}} C_\Gamma 
\end{align*}

\subsubsection{Interior stencils of $\cSub$}
For constructing the interior stencils of $\cSub$, we assume that the stencil coefficients are mirror-symmetric with respect to the subdivided edge element. In addition, as in~\cite{Wang:2006}, we fix the odd stencil for $\cSub$ with the same global shape as the $S_1$ odd stencil. Finally, we demand that the coefficients for even stencils of valence $\geq 7$ are the same over the finite support of $S_\mathcal{F^{\ast}}$.

After construction of the new $S_\mathcal{E^*}$ operator via the commutations, there are three degrees of freedom remaining. We resolve two of them by requiring all coefficients of the even valence 6 stencil to be positive. The remaining degree of freedom is present in the valence 4 even stencil, for which the local subdivision operator spectrum is $[1/4, 3/16, 3/16, 1/8, 1/8, 1/8, 1/16, 3/16 - 4 z]$, where $z$ is the remaining degree of freedom. We choose $z=1/32$ to make the spectrum consist of $1\times1/4, 2\times,3/16,3\times1/8,2\times1/16$. 

\begin{figure}
%\begin{minipage}{0.45\textwidth}
\includegraphics[width = 0.5\textwidth]{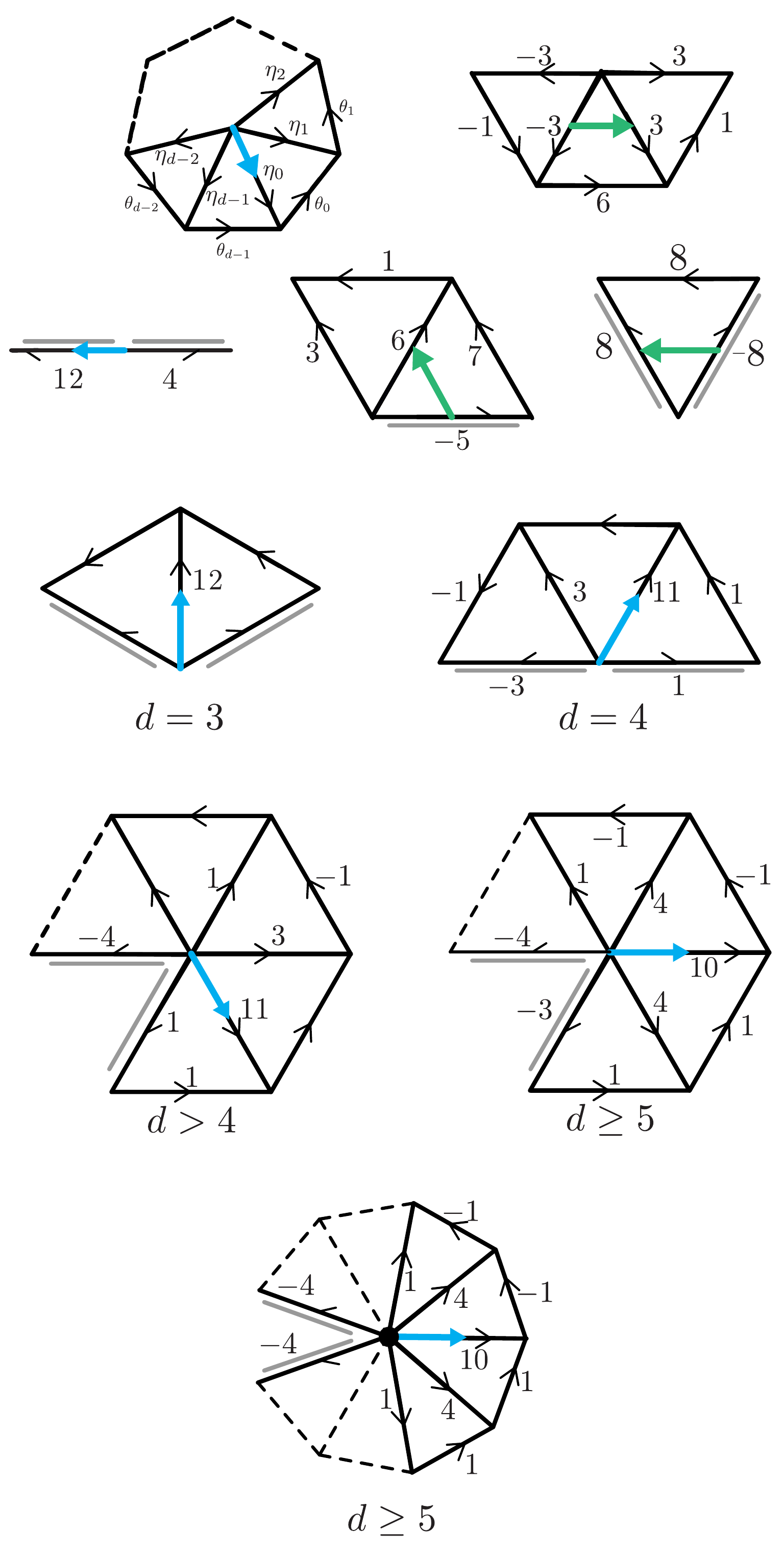}
\caption{$1$-form subdivision operator $S_1$. Arrows denote assumed edge direction. Multiply all factors by $1/32$ except for the $\eta_i,\theta_i$ factors. In the bottommost stencil, when the boundary edge and the first edge of the interior stencil coincide, the coefficients should be summed up together. Similar goes for the odd interior stencil for valence 3.}
\label{fig:S1}
\end{figure}

The stencils of the $\cSub$ operator are summarized in Figure~\ref{fig:Se}. The coefficients $\zeta$ and $\zeta'$ for the valence 5 interior rings are given by
\begin{equation}
% \zeta = \frac{1}{16}\frac{1}{\sqrt{5} + 5}	
\zeta = \frac{2}{\sqrt{5} + 5}, \zeta' = 1-\zeta
\end{equation}

\subsection{Boundary stencils}
\begin{figure}
\includegraphics[width = 0.5\textwidth]{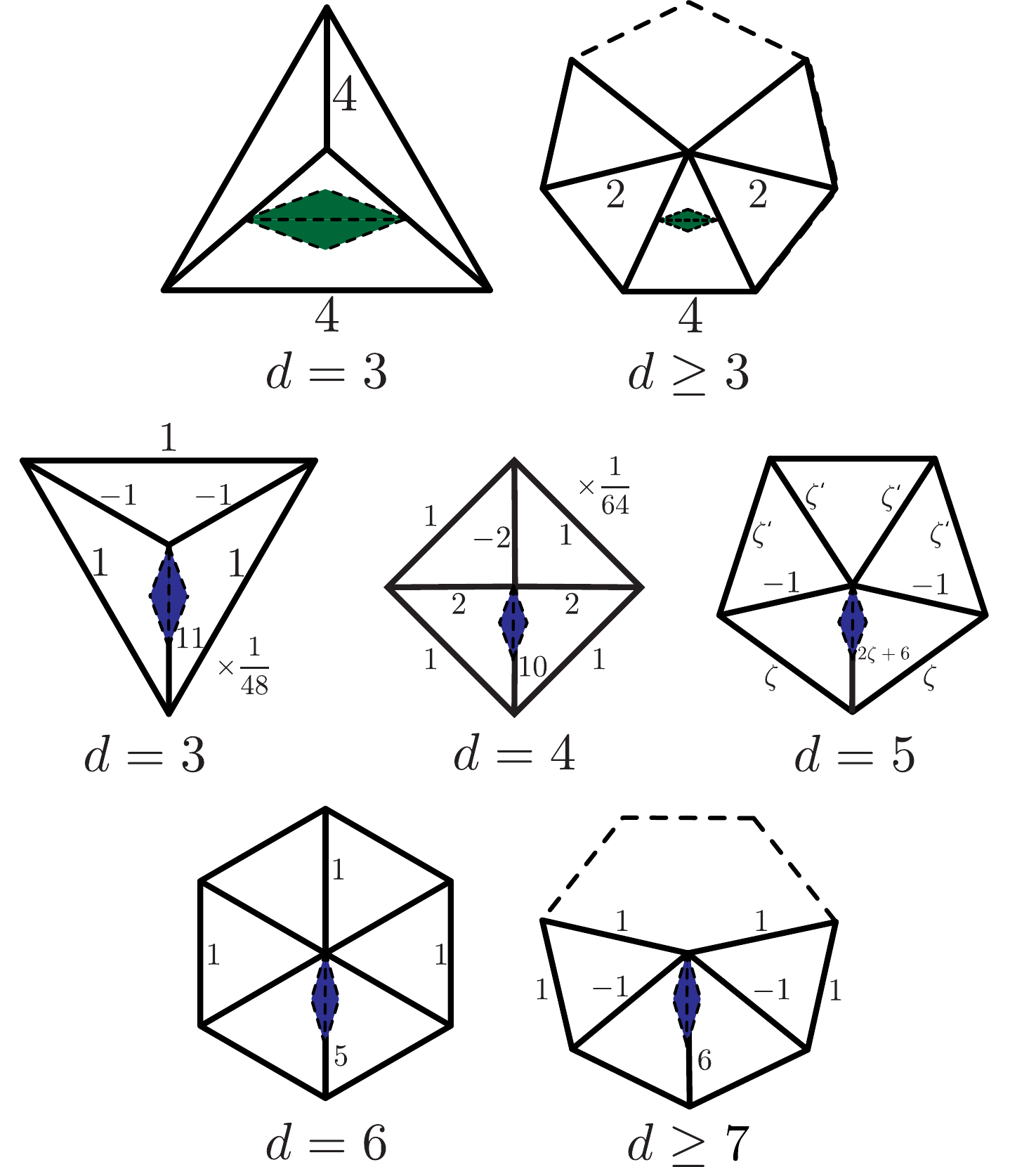}\\
\includegraphics[width = 0.5\textwidth]{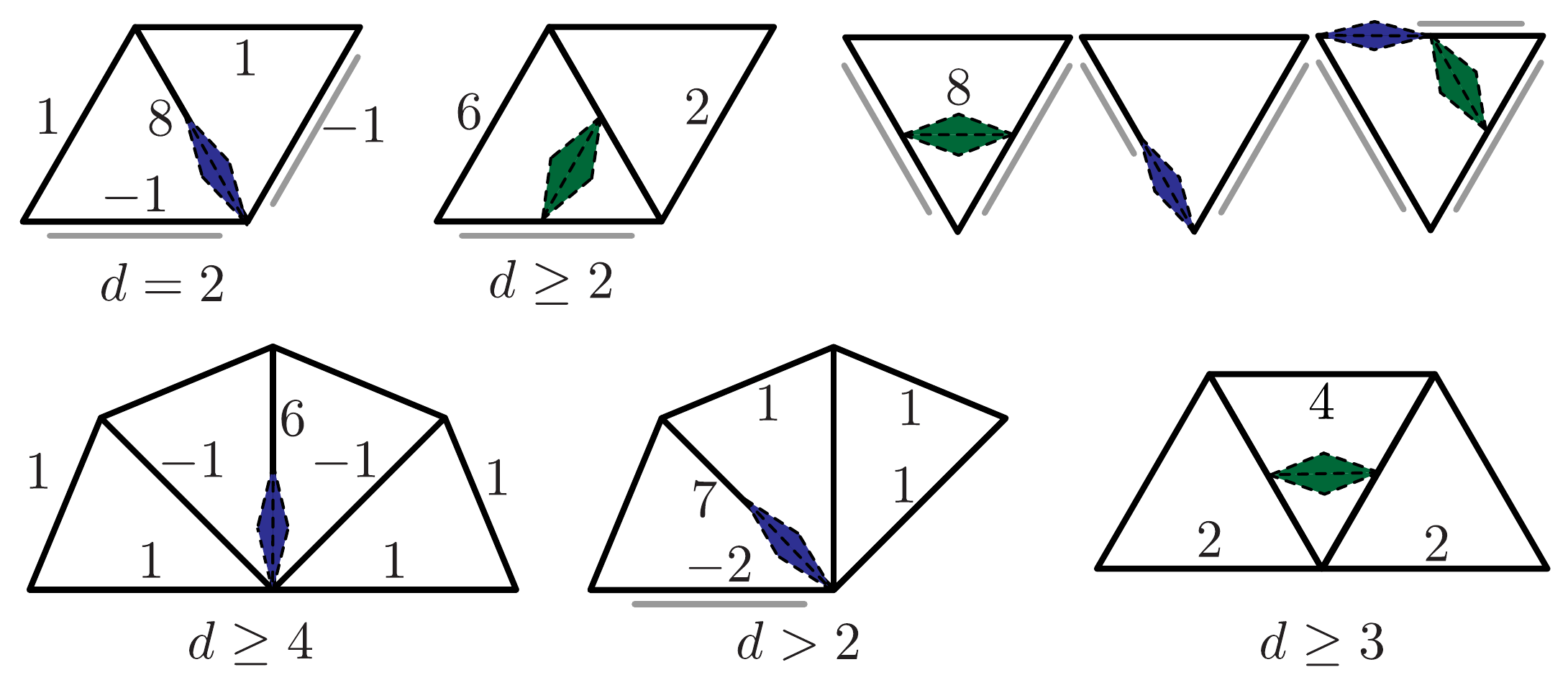}
\caption{Unsigned integrated edge subdivision $S_{\mathcal{E}^{\ast}}$. Multiply all factors by $1/32$ unless stated otherwise. Double edges denote boundary edges. When the outer edges with coefficients of the interior odd stencil coincide for valence 3, sum the coefficients.}
\label{fig:Se}
%\end{minipage}
\end{figure}

We assume all boundary stencils to be applied with mirror symmetry around the boundary. Since we want to preserve the $C\gamma = 0$ condition on the boundary, we demand that the coefficients for the boundary stencil for even elements of $\cSub$ solely depend on the boundary. In addition, we require that the odd stencil for the elements that touch the boundary with one of their vertices remains the same for $d\geq3$. Using these assumptions on the stencils, we solve for $\cSub$, $\hbSub$ and $\formSub$ in conjunction. The resulting subdivision operators have a single degree of freedom left, which we resolve by requiring all $\hbSub$ elements to be positive.  The modified stencils for $S_{\mathcal{F}^{\ast}}$ are shown in Figure~\ref{fig:Sf}, for $S_1$ in Figure~\ref{fig:S1}, and for $S_{\mathcal{E}^{\ast}}$ in Figure~\ref{fig:Se}.

%\end{minipage}
%\begin{minipage}{0.45\textwidth}

\end{document}